\documentclass[]{jfm}
\usepackage{graphicx}
\usepackage{epstopdf, epsfig}
\usepackage{amsmath, amssymb}
\usepackage{bm}
\usepackage{here}
\usepackage{graphicx,siunitx}
\usepackage{listings,color,courier,natbib}
\usepackage{pdflscape, csquotes}
\graphicspath{{./figures/}}
\usepackage[abs]{overpic}

\usepackage[T1]{fontenc}
\usepackage{adjustbox}
\usepackage{rotating}
\usepackage{algorithm}
\usepackage{algorithmic}
\usepackage{lscape}
\usepackage{amsfonts}
\usepackage{xurl}
\usepackage{color}
\usepackage{natbib}
\definecolor{green}{rgb}{0.3, 0.45, 0.02}

\shorttitle{SINDy with low-dimensionalized flow representations}
\shortauthor{K. Fukami, T. Murata, K. Zhang, and K. Fukagata}

\title{Sparse identification of nonlinear dynamics with low-dimensionalized flow representations}

\author{Kai Fukami\aff{1,2}
  \corresp{\email{kfukami1@g.ucla.edu}},
  Takaaki Murata\aff{2},
  Kai Zhang\aff{3},
 \and Koji Fukagata\aff{2}}

\affiliation{\aff{1}Department of Mechanical and Aerospace Engineering, University of California, Los Angeles, CA 90095
\aff{2}Department of Mechanical Engineering, Keio University, Yokohama, 223-8522, Japan
\aff{3}Department of Mechanical and Aerospace Engineering, Rutgers University, Piscataway, NJ 08854, USA
}

\begin{document}

\maketitle

\begin{abstract}
{We perform a sparse identification of nonlinear dynamics (SINDy) for low-dimensionalized complex flow phenomena.
We first apply the SINDy with two regression methods, the thresholded least square algorithm (TLSA) and the adaptive Lasso (Alasso) which show reasonable ability with a wide range of sparsity constant in our preliminary tests, to a two-dimensional single cylinder wake at $Re_D=100$, its transient process, and a wake of two-parallel cylinders, as examples of high-dimensional fluid data.
To handle these high dimensional data with SINDy whose library matrix is suitable for low-dimensional variable combinations, a convolutional neural network-based autoencoder (CNN-AE) is utilized.
The CNN-AE is employed to map a high-dimensional dynamics into a low-dimensional latent space.
The SINDy then seeks a governing equation of the mapped low-dimensional latent vector.
Temporal evolution of high-dimensional dynamics can be provided by combining the predicted latent vector by SINDy with the CNN decoder which can remap the low-dimensional latent vector to the original dimension.
The SINDy can provide a stable solution as the governing equation of the latent dynamics and the CNN-SINDy based modeling can reproduce high-dimensional flow fields successfully, although more terms are required to represent the transient flow and the two-parallel cylinder wake than the periodic shedding.
A nine-equation turbulent shear flow model is finally considered to examine the applicability of SINDy to turbulence, although without using CNN-AE.
The present results suggest that the proposed scheme with an appropriate parameter choice enables us to analyze high-dimensional nonlinear dynamics with interpretable low-dimensional manifolds.}

\end{abstract}

\begin{keywords}
computational methods, machine learning, low-dimensional models
\end{keywords}

\vspace{-15mm}
\section{Introduction}

{
Sparse identification of nonlinear dynamics (SINDy)~\citep{BPK2016a} is one of the prominent data-driven tools to obtain governing equations of nonlinear dynamics in a form that we can understand.
The SINDy algorithm enables us to discover a governing equation from a time-discretized data and identify dominant terms from a large set of potential terms that are likely to be involved in the model.
Recently, the usefulness of the SINDy has been demonstrated in various fields~\citep{CBK2019,HFN2019,ZS2019,DNMP2020}.
Here, let us introduce some efforts, especially in fluid dynamics community.
\citet{LNB2018}~utilized SINDy to present general reduced order modelling (ROM) framework for experimental data: sensor-data and particle image velocimetry data.
The model was investigated using a transient and post-transient laminar cylinder wake.
They reported that the nonlinear full-state dynamics can be modeled with sensor-based dynamics and SINDy-based estimation for coefficients of ROM.
The SINDy with taking into account control inputs (called SINDYc) was also investigated by~\citet{BPK2016b} using the Lorenz equations.
\citet{LB2018}~combined SINDy and POD to enforce energy-preserving physical constraints in the regression procedure toward the development of a new data-driven Galerkin regression framework.
For the time-varying aerodynamics,~\citet{LKLLBK2019} identified vortex-induced vibrations on a long-span suspension bridge utilizing the SINDy algorithm extended to parametric partial differential equations \citep{RBPK2017}.
As a novel method to perform the order reduction of data and SINDy simultaneously, there is a customized autoencoder~\citep{CLKB2019} introducing the SINDy loss in the
loss function of deep autoencoder networks.
The sparse regression idea has also recently been propagated to turbulence closure modeling purposes~\citep{schmelzer2020discovery,beetham2020formulating,beetham2021sparse,duraisamy2020machine}.
As reported in these studies, by employing the SINDy to predict the temporal evolution of a system, we can obtain ordinary differential equations, which should be helpful to many applications, e.g., control of a system.
In this way, the propagation of the use of SINDy can be seen in the fluid dynamics community.

\begin{figure}
    \begin{center}
        \includegraphics[width=0.80\textwidth]{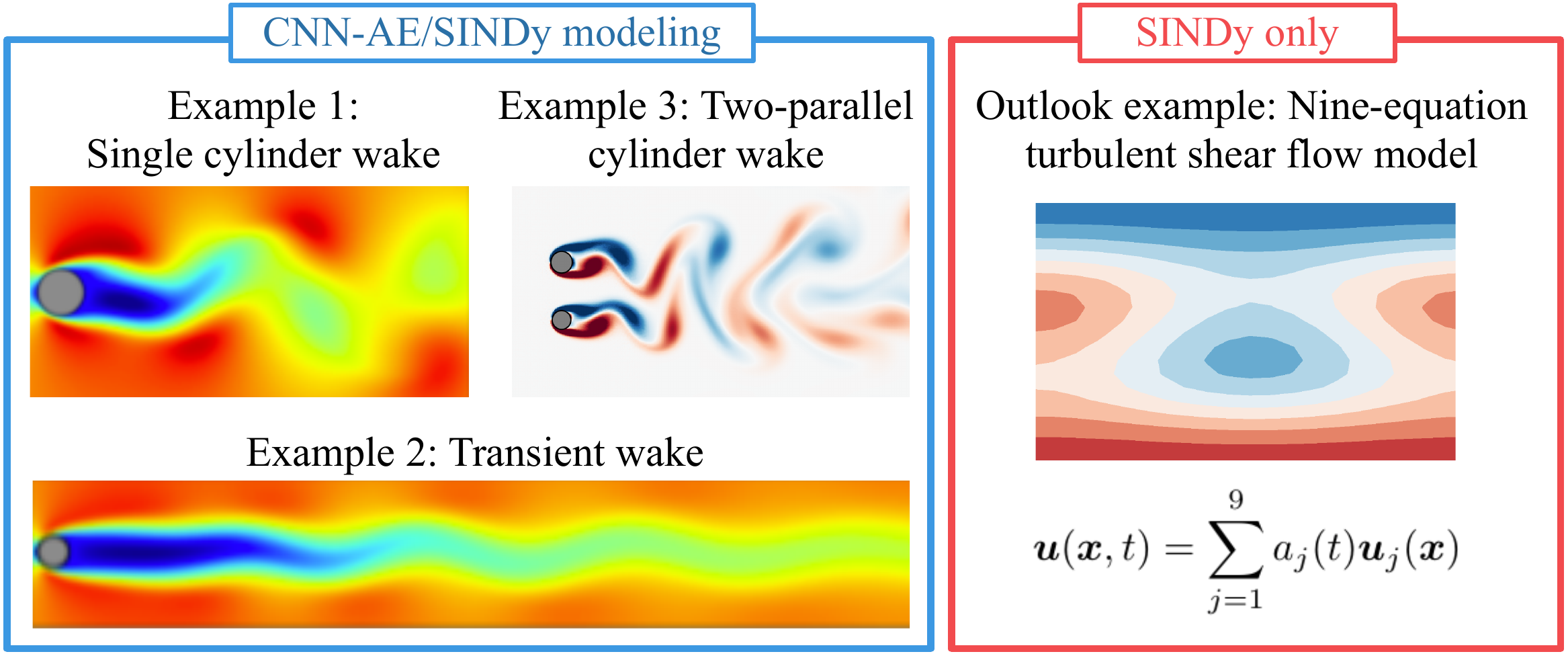}
    \caption{Covered examples of fluid flows in the present study.}
    \label{fig1}
    \end{center}
\end{figure}

We here examine a possibility of SINDy-based modeling of low-dimensionalized complex fluid flows presented in figure~\ref{fig1}.
Following our preliminary tests with the van der Pol oscillator and the Lorenz attractor presented in Appendix, we apply the SINDys with two regression methods, 1. TLSA and 2. Alasso, to a two-dimensional cylinder wake at $Re_D=100$, its transient process, and a wake of two-parallel cylinders as examples of high-dimensional dynamics.
In this study, a convolutional neural network-based autoencoder (CNN-AE) is utilized to handle these high dimensional data with SINDy whose library matrix is suitable for low-dimensional variable combinations~\citep{KKB2018}.
The CNN-AE here is employed to map a high-dimensional dynamics into low-dimensional latent space.
The SINDy is then utilized to obtain a governing equation of the mapped low-dimensional latent vector.
Unifying the dynamics of the latent vector obtained via SINDy with the CNN decoder which can remap the low-dimensional latent vector to the original dimension, we can present the temporal evolution of high-dimensional dynamics and avoid the issue with high-dimensional variables as discussed above.
The scheme of the present ROM is inspired by~\citet{HFMF2019AJK,HFMF2020b,HFMF2019} who utilized a long short term memory instead of SINDy in the present ROM framework.
The paper is organized as follows: we provide details of SINDy and CNN-AE in section~\ref{sec:methods}.
We present the results for high-dimensional flow examples with the CNN-AE/SINDy reduced-order modeling in sections~\ref{pcw}, \ref{tcw} and \ref{sec:tpcw}.
In section~\ref{sec:nine-eq}, we also discuss its applicability to turbulence using the nine-equation turbulent shear flow model, although without considering the CNN-AE.
At last, concluding remarks with outlook are offered in section~\ref{sec:conc}.
}

\section{Methods}
\label{sec:methods}

{

\subsection{Sparse identification of nonlinear dynamics (SINDy)}

Sparse Identification of Nonlinear Dynamics (SINDy) \citep{BPK2016a} is performed to identify nonlinear governing equations from time series data in the present study.
The temporal evolution of the state $\bm{x}(t)$ in a typical dynamical system can be often represented in a form of ordinary differential equation,
\begin{align}
    \dot{\bm{x}}(t) = \bm{f} ( \bm{x}(t) ). \label{eq:state_eq}
\end{align}
To explain SINDy algorithm, let $\bm{x}(t)$} be $(x(t),y(t))$ hereinafter, although the SINDy can also handle the dynamics with higher dimensions as will be considered later. 
First, the temporally discretized data of $\bm{x}$ are collected to arrange a data matrix $\bm{X}$,
\begin{align}
  \bm{X}=
  \left( 
    \begin{array}{c}
      \bm{x}^T(t_1)  \\
      \bm{x}^T(t_2)  \\
       \vdots  \\
      \bm{x}^T(t_m) \\
    \end{array}
     \right)
    =
  \left( 
    \begin{array}{cc}
      x(t_1) & y(t_1) \\
      x(t_2) & y(t_2) \\
       \vdots  &  \vdots  \\
      x(t_m) & y(t_m) \\
    \end{array}
     \right).
\end{align}
We then collect the time series data of the time-differentiated value $\dot{\bm x}(t)$ to construct a time-differentiated data matrix $\dot{\bm{X}}$,
\begin{align}
    \dot{\bm{X}}= 
    \left( 
    \begin{array}{c}
      \dot{\bm{x}}^T(t_1)  \\
      \dot{\bm{x}}^T(t_2)  \\
       \vdots  \\
      \dot{\bm{x}}^T(t_m) \\
    \end{array}
     \right)
    =
    \left( 
      \begin{array}{cc}
        \dot{x}(t_1) & \dot{y}(t_1) \\
        \dot{x}(t_2) & \dot{y}(t_2) \\
         \vdots  &  \vdots  \\
        \dot{x}(t_m) & \dot{y}(t_m) \\
      \end{array}
    \right).
\end{align}
In the present study, the time-differentiated values are obtained with the second-order central-difference scheme.
Note in passing that results in the present study are not sensitive to the differential scheme with appropriate time stpdf for construction of $\dot{\bm X}$.
Then, we prepare a library matrix $\Theta (\bm{X})$ including nonlinear terms of $\bm{X}$,
\begin{align}
    \Theta (\bm{X}) = \left( 
      \begin{array}{cccccc}
         | &   |    &   |    &   |      &   |    &   | \\
         1 & \bm{X} & \bm{X}^{P_2} & \bm{X}^{P_3} &\bm{X}^{P_4} &\bm{X}^{P_5} \\
         | &   |    &   |    &   |      &   |    &   | \\
      \end{array}
    \right), \label{eq:library}
\end{align}
where $\bm{X}^{P_i}$ is $i$th order polynomials constructed by $x$ and $y$.
The set of nonlinear potential terms here includes up to 5th order terms, although what terms are included can be arbitrarily set.
We finally determine a coefficient matrix $\Xi$,
\begin{align}
    \Xi = (\xi_x ~~ \xi_y) = \left( 
      \begin{array}{cc}
        \xi _{(x,~1)} & \xi _{(y,~1)} \\
        \xi _{(x,~2)} & \xi _{(y,~2)} \\
        \vdots & \vdots \\
        \xi _{(x,~l)} & \xi _{(y,~l)} \\
      \end{array}
    \right),
    \label{eq:2.5}
\end{align}
in the state equation,
\begin{align}
    \dot{\bm{X}}(t)=\Theta (\bm{X}) \Xi, \label{eq:sindy}
\end{align}
using an arbitrary regression method, such as TLSA, Lasso, and so on. 
The subscript $l$ in equation (\ref{eq:2.5}) denotes the row index
of the library matrix.
Once the coefficient matrix $\Xi$ is obtained, the governing equation is identified as
\begin{align}
    \dot{x}=\Theta(x) \xi_x,~~\dot{y}= \Theta(y) \xi_y.
\end{align}

In this study, we use the thresholded least square algorithm (TLSA) used in~\citet{BPK2016a} and the adaptive Lasso (Alasso)~\citep{Zou2006} to obtain the coefficient matrix $\Xi$ following our preliminary tests. 
Details can be found in Appendix.

\subsection{Convolutional neural network-based autoencoder}

\begin{figure}
    \begin{center}
        \includegraphics[width=0.85\textwidth]{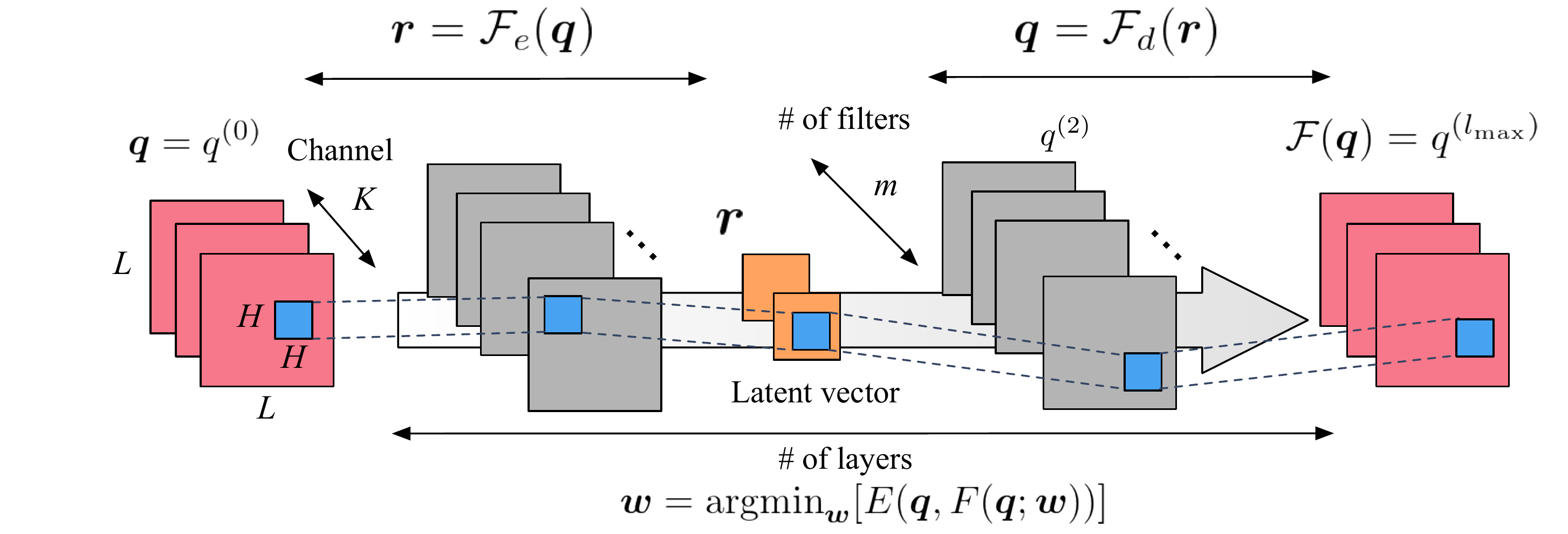}
    \caption{Convolutional neural network-based autoencoder.}
    \label{fig_f1}
    \end{center}
\end{figure}

We use a convolutional neural network \citep{LBBH1998}-based autoencoder \citep{HS2006} (CNN-AE) for order reduction of high-dimensional data, as shown in figure \ref{fig_f1}.
As an example, we show the CNN-AE with three hidden layers.
In a training process, the autoencoder $\cal F$ is trained to output the same data as the input data $\bm q$ such that ${\bm q}\approx{\cal F}({\bm q};{\bm w})$, where ${\bm w}$ denotes the 
weights of the machine learning model.  
The process to optimize the weights $\bm w$ can be formulated as an iterative
minimization of an error function $E$,
\begin{eqnarray}
{\bm w}={\rm argmin}_{\bm w}[{E}({\bm q},{\mathcal F}({\bm q};{\bm w}))].
\end{eqnarray}
For the use of autoencoder as a dimension compressor, the dimension of an intermediate space called the latent space $\bm \eta$ is smaller than that of the input or output data $\bm q$ as illustrated in figure \ref{fig_f1}. 
When we are able to obtain the output ${\cal F}(\bm q)$ similar to the input $\bm q$ such that ${\bm q}\approx{\cal F}({\bm q})$, it can be guaranteed that the latent vector $\bm r$ is a low-dimensional representation of its input or output $\bm q$.
In an autoencoder, the dimension compressor is called as the encoder ${\cal F}_e$ (the left part in figure \ref{fig_f1}) and the counterpart is referred to as the decoder ${\cal F}_d$ (the right part in figure \ref{fig_f1}).
Using them, the internal procedure of the autoencoder can be expressed as
\begin{eqnarray}
{\bm{r}} ={\mathcal F}_{e}({\bm q}), ~~{\bm{q}} ={\mathcal F}_{d}(\bm r).
\end{eqnarray}
In the present study, the dimension of the latent space for the problems of a cylinder wake and its transient process is set to 2 following our previous work~\citep{MFF2020}.
\citet{MFF2020} reported that the flow fields can be mapped into a low-dimensional latent space successfully while keeping the information of high-dimensional flow fields. 
For the example of a two-parallel cylinder wake in section~\ref{sec:tpcw}, the dimension of the latent space is set to 4 to handle its quasi-periodicity accordingly.
For construction of the autoencoder, the $L_2$ norm error is applied as the error function $E$ and the Adam optimizer~\citep{kingma2014} is utilized for updating the weights in the iterative training process.

\begin{figure}
    \begin{center}
        \includegraphics[width=0.8\textwidth]{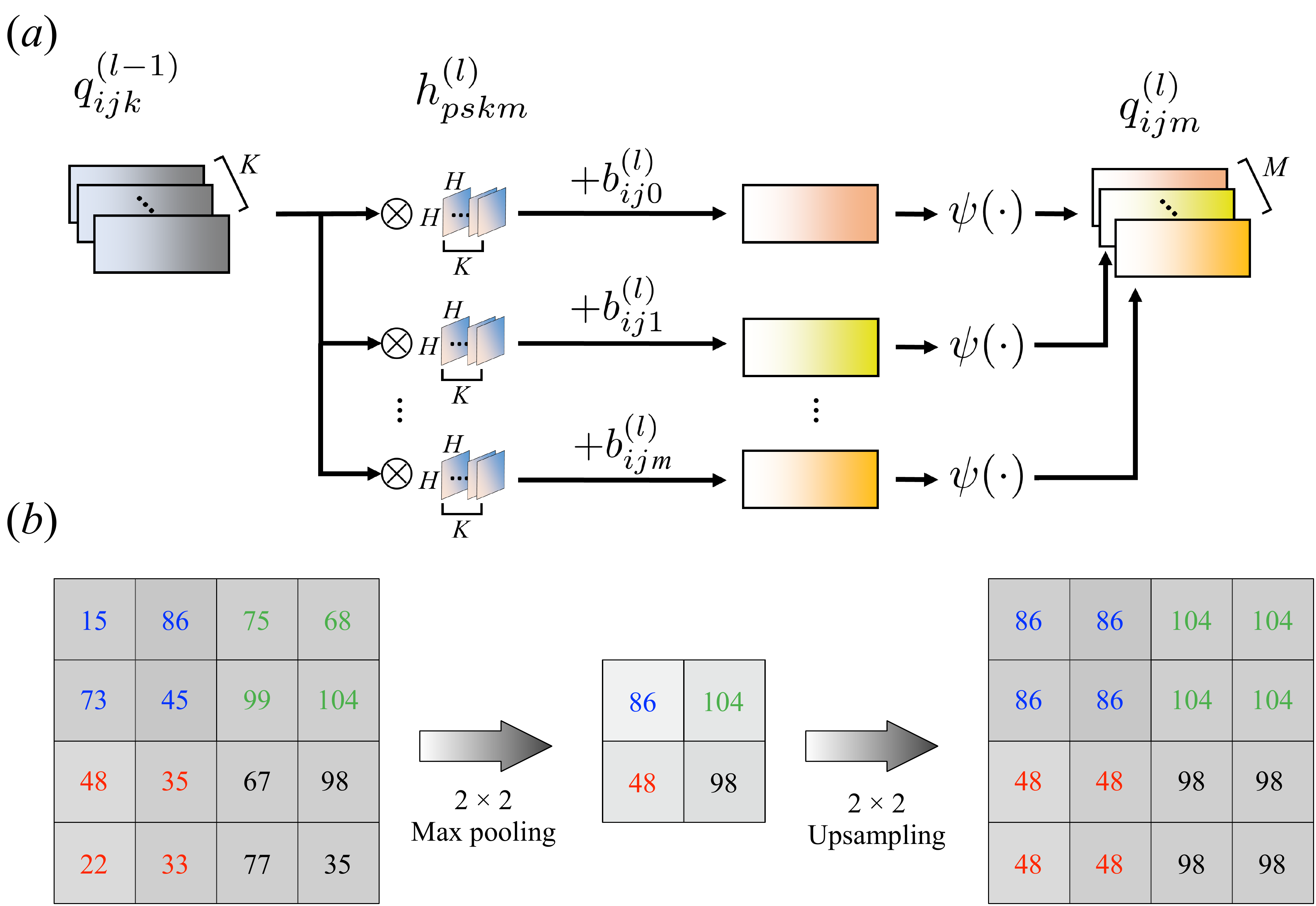}
    \caption{Internal procedure of convolutional neural network. $(a)$ Filter operation with activation function. $(b)$ Max pooling and upsampling.}
    \label{fig_f2}
    \end{center}
\end{figure}

Next, let us briefly explain the convolutional neural network.
In the scheme of convolutional neural network, a filter $h$, whose size is $H\times H,$ is utilized as the weights $\bm w$ as shown in figure \ref{fig_f2}$(a)$.
The mathematical expression here is 
\begin{eqnarray}
    q^{(l)}_{ijm} = {\psi}\biggl(\sum^{K-1}_{k=0}\sum^{H-1}_{p=0}\sum^{H-1}_{s=0}h_{p{s}km}^{(l)} q_{i+p-C,j+{s-C},k}^{(l-1)} + b^{(l)}_{ijm}\biggr),
\end{eqnarray}
where {$C={\rm floor}(H/2)$,} $q^{(l)}$ is the output at layer $l$, and $K$ is the number of variables in the data (e.g., $K=1$ for black-and-white photo and $K=3$ for RGB images).
Although not shown in figure \ref{fig_f2}, $b$ is a bias added to the results of filter operation. 
The activation function $\psi$ is usually a monotonically increasing nonlinear function.
The autoencoder can achieve more effective dimension reduction than the linear theory-based method, i.e., proper orthogonal decomposition (POD)~\citep{Lumely1967}, thanks to the nonlinear activation function~\citep{Milano2002,MFF2020,FHNMF2020}. 
In the present study, a hyperbolic tangent function $\psi(a)=(e^{a}-e^{-a})\cdot(e^{a}+e^{-a})^{-1}$ is adopted for the activation function following~\citet{MFF2020}. 
Moreover, we use the convolutional neural network to construct the autoencoder because the CNN is good at handling high-dimensional data with lower computational costs than fully-connected models, i.e., multi-layer perceptron, thanks to the filter concept which assumes that pixels of images have no strong relationship with those of far areas.
Recently, the use of CNN has been emerged to deal with high dimensional problems including fluid dynamics~\citep{FFT2019b,FFT2019a,FNKF2019,omata2019,FFT2020b,MFF2020b,morimoto2101convolutional,FFT2019tsfp,FukamiVoronoi,matsuo2021supervised}, although this concept was originally developed in the field of computer science.

\subsection{CNN-SINDy based reduced-order modeling}

\begin{figure}
    \begin{center}
        \includegraphics[width=0.70\textwidth]{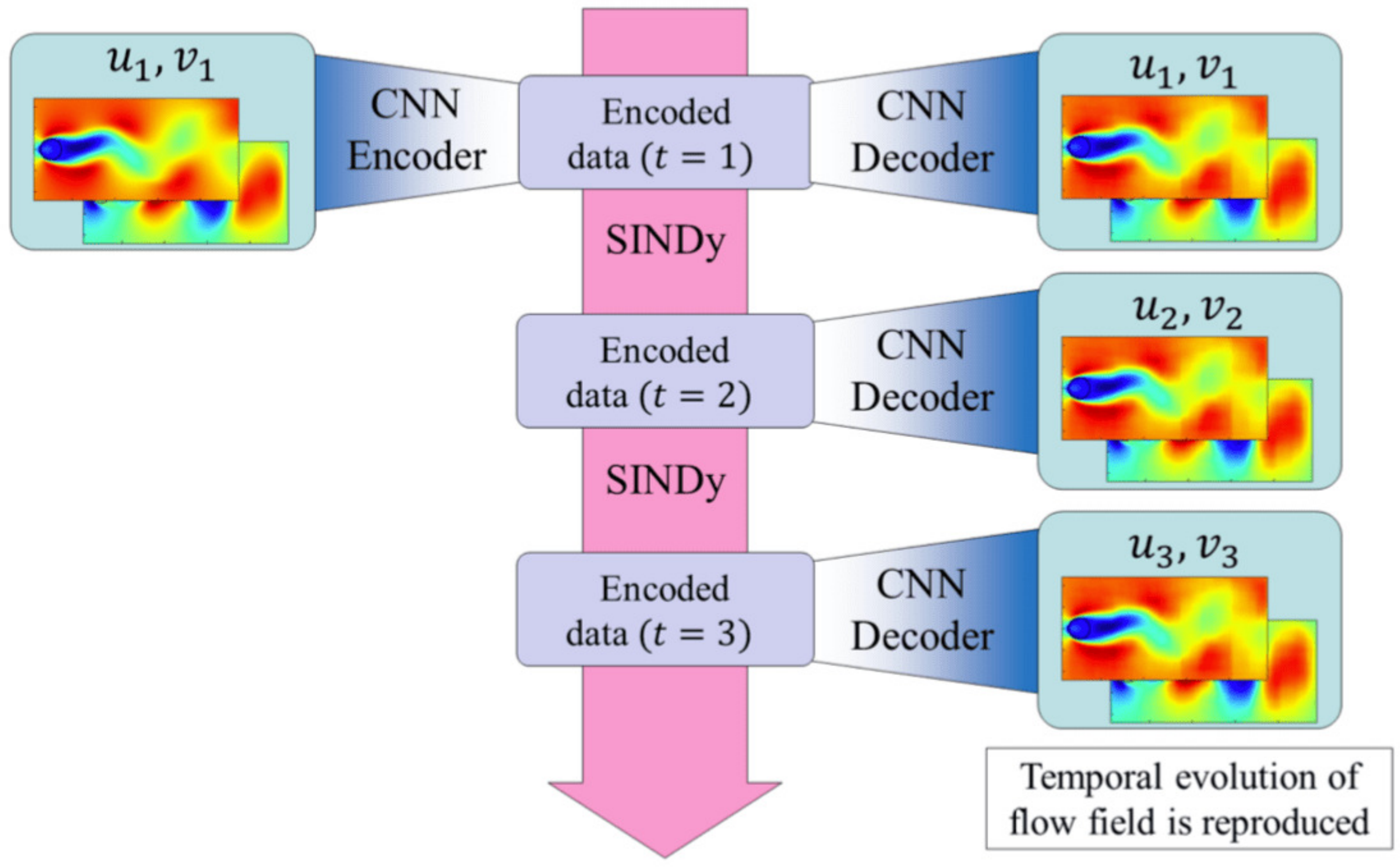}
    \caption{CNN-SINDy based reduced-order modeling for fluid flows.}
    \label{fig_rom}
    \end{center}
\end{figure}

By combining the CNN-AE and SINDy, we present the machine-learning-based reduced-order modeling (ML-ROM) as shown in figure~\ref{fig_rom}.
The CNN-AE first works to map a high-dimensional flow field into low-dimensional latent space.
The SINDy is then performed to obtain a governing equation of the mapped low-dimensional latent vector.
Unifying the predicted latent vector by the SINDy with the CNN decoder, we can obtain the temporal evolution of high-dimensional flow field as presented in figure~\ref{fig_rom}.
Moreover, the issue with high-dimensional variables and SINDy can also be avoided.
Note again that the present ROM is inspired by~\citet{HFMF2019AJK,HFMF2020b,HFMF2019} who capitalized on a long short term memory instead of SINDy.
We apply this framework to a two-dimensional cylinder wake in section~\ref{pcw}, its transient process~in section~\ref{tcw}, and a wake of two-parallel cylinders in section~\ref{sec:tpcw}.
In this study, we perform a five-fold cross-validation~\citep{Bruntonkutz2019} to create all machine learning models, although only the results of a single case will be presented for brevity.
The sample code for the present reduced-order model is available from \url{https://github.com/kfukami/CNN-SINDy-MLROM}.

\section{Results and discussion}
\subsection{Example 1: periodic cylinder wake}
\label{pcw}

Here, let us consider a two-dimensional cylinder wake using a two-dimensional direct numerical simulation (DNS).
The governing equations are the incompressible continuity and
Navier--Stokes equations,
\begin{align}
    &\bm{\nabla} \cdot \bm{u}=0, \\
    &\dfrac{\partial\bm{u}}{\partial t} + \bm{\nabla} \cdot (\bm{uu}) =  - \bm{\nabla} p + \frac{1}{{Re}_D}\nabla ^2 \bm{u},
\end{align}
where $\bm{u}$ and $p$ represent the velocity vector and pressure, respectively.
All quantities are made dimensionless by the fluid density, the free-stream velocity and the cylinder diameter.
The Reynolds number based on the cylinder diameter is $Re _D=100$.
The size of the computational domain is $L_x=25.6$ and $L_y=20.0$ in the streamwise ($x$) and the transverse ($y$) directions, respectively.
The origin of coordinates is defined at the center of the inflow boundary.
A Cartesian grid with the uniform grid spacing of $\Delta x=\Delta y = 0.025$ is used; thus, the number of grid points is $(N_x, N_y)=(1024, 800)$.
We use the ghost cell method \citep{kor2017} to impose the no-slip boundary condition on the surface of cylinder whose centre is located at $(x,y)=(9,0)$.
In the present study, we utilize the flows around the cylinder as the training data set, i.e., $8.2 \leq x \leq 17.8$ and $-2.4 \leq y \leq 2.4$ with $(N_x^*, N_y^*)=(384, 192)$.
The fluctuation components of streamwise velocity $u$ and transverse velocity $v$ are considered as the input and output attributes for CNN-AE. 
The time interval of the flow field data is $\Delta t = 0.025$ corresponding to approximately 230 snapshots per a period with the Strouhal number $St=0.172$.

\begin{figure}
  \centerline{\includegraphics[clip,width=0.7\linewidth]{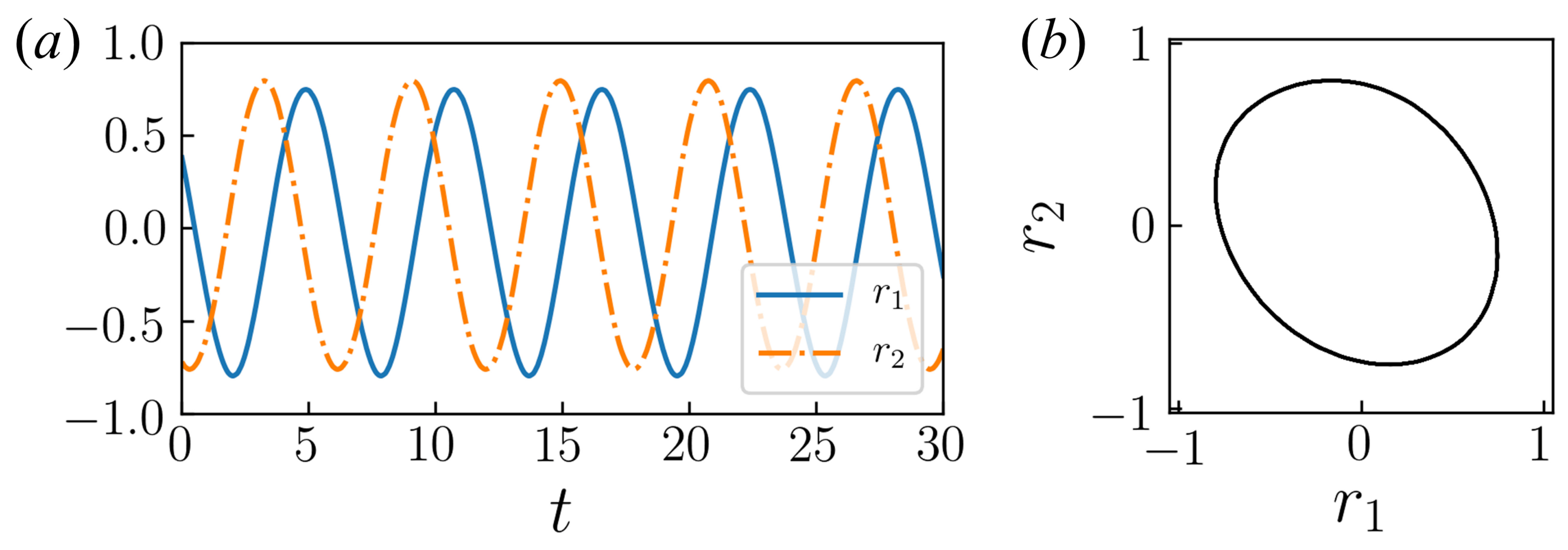}}
  \caption{Latent vector dynamics of a periodic shedding case: $(a)$ time history and $(b)$ trajectory.}
  \label{fig_per}
\end{figure}
As mentioned above, the SINDy is employed to identify the ordinary differential equations that govern the temporal evolution of the mapped vector 
obtained by the CNN encoder.
The time history and the trajectory of the mapped latent vector ${\bm r}=(r_1,r_2)$ are shown in figure~\ref{fig_per}.
For the assessment of the candidate models, we integrate the differential equations to reproduce the time history.
Note in passing that the indices based on $L_2$ error are not suitable since the slight difference between the period of reproduced oscillation and that of original one results in large $L_2$ errors for oscillating systems.
We here use the amplitude and the frequency of oscillation to evaluate the similarity between the reproduced waveform and the original one.
Hereafter, the error rate of the amplitude and the frequency are shown in the figures below for simplicity.
The number of terms is also considered for the parsimonious model selection.

\begin{figure}
  \centerline{\includegraphics[clip,width=1.0\linewidth]{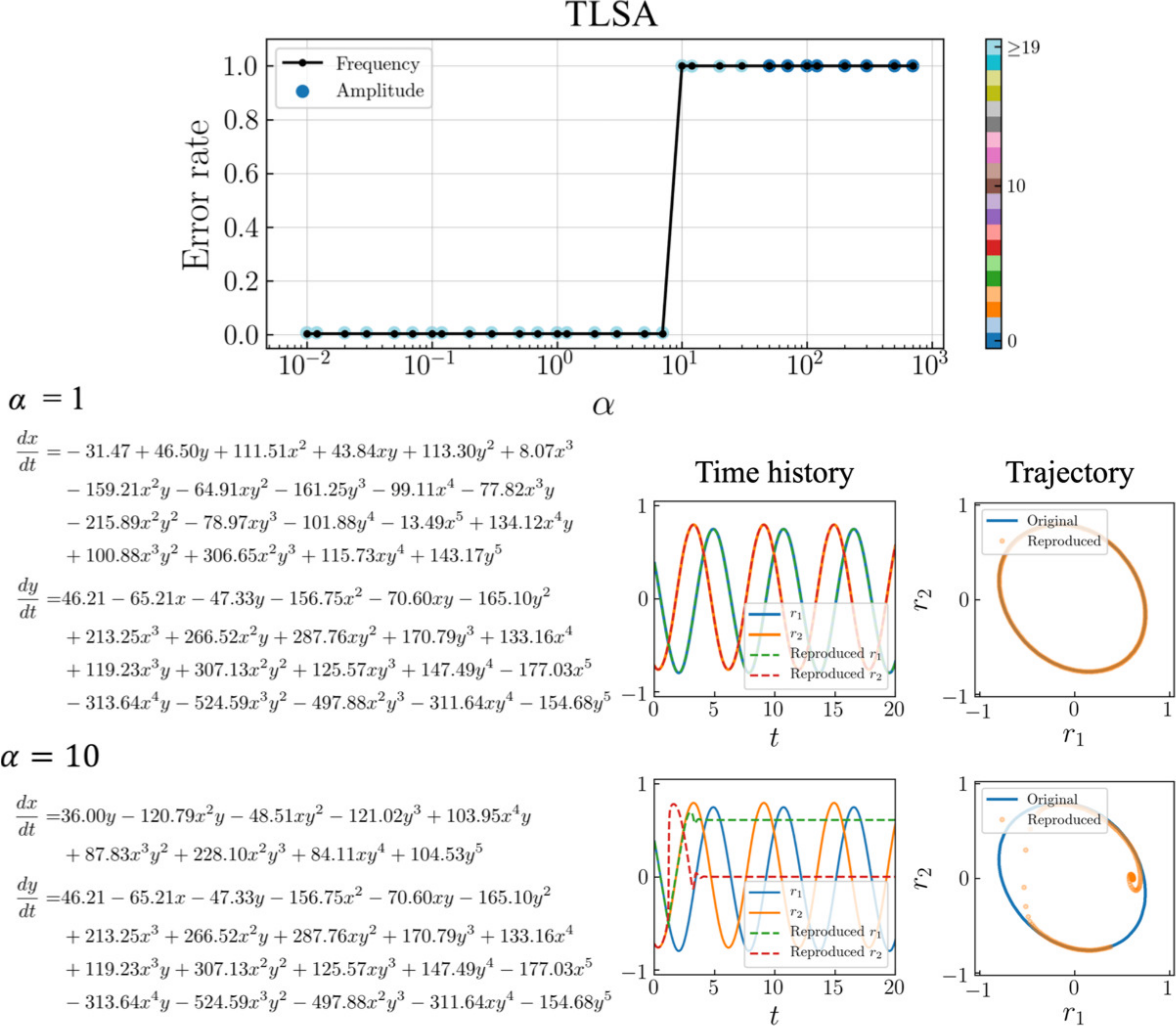}}
  \caption{Results with TLSA for the periodic cylinder example.
  Parameter search results, examples of obtained equations, and reproduced trajectories with $\alpha=1$ and $\alpha=10$ are shown. Color of amplitude plot indicates the total number of terms.  In the ordinary differential equations, the latent vector components $(r_1, r_2)$ are represented by $(x,y)$ for the distinctness.}
  \label{fig_per_tlsa}
\end{figure}
\begin{figure}
  \centerline{\includegraphics[clip,width=0.75\linewidth]{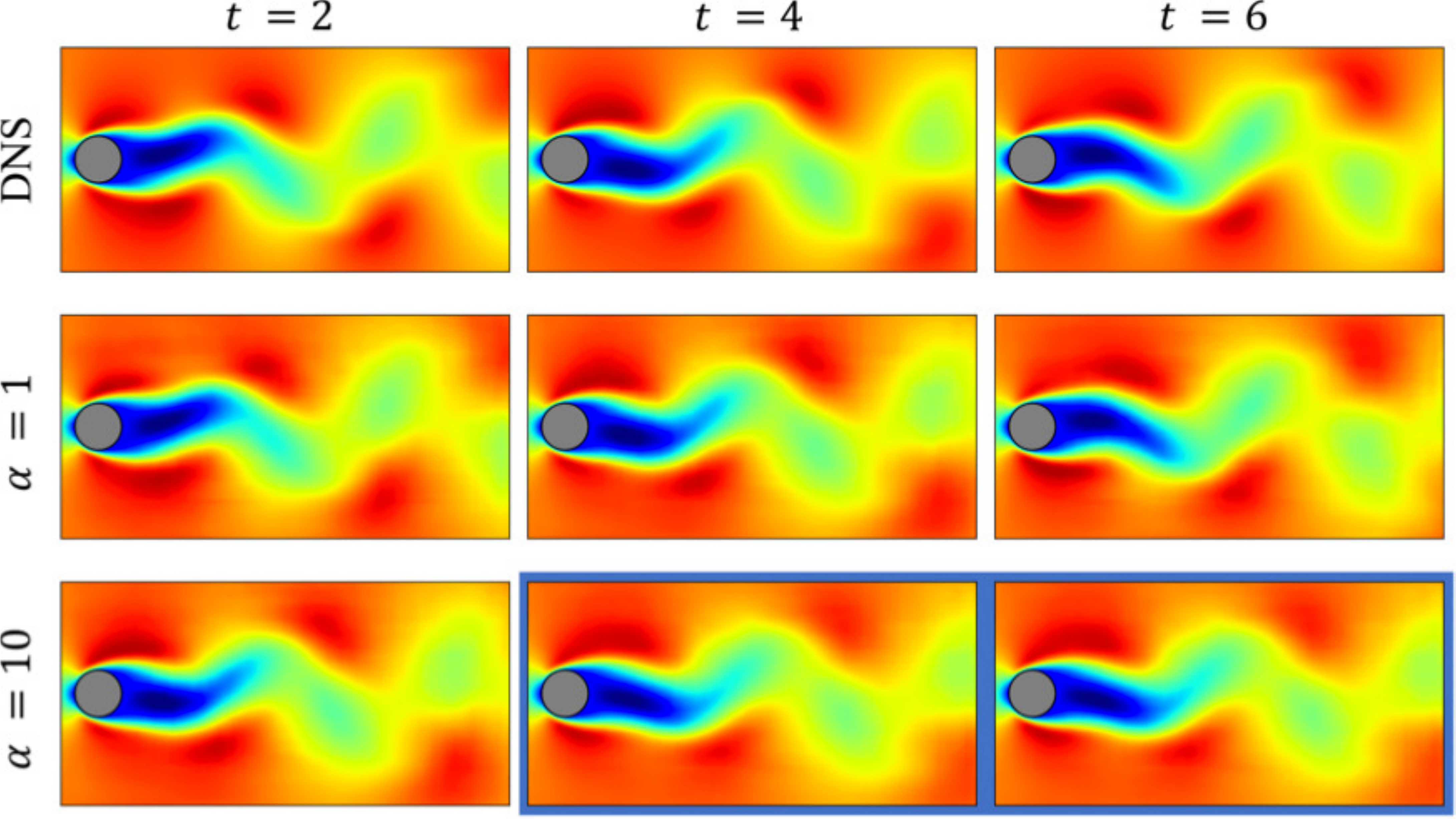}}
  \caption{Temporally evolved flow fields of DNS and the reproduced flow field with $\alpha=1$ and $\alpha=10$. With $\alpha=10$, the flow field 
  is frozen after around $t=4$ (enhanced using blue portion).}
  \label{fig_per_tlsa_field}
\end{figure}

We consider two regression methods for SINDy, i.e., TLSA and Alasso, following our previous tests.
Let us present the results of parameter search utilizing TLSA in figure~\ref{fig_per_tlsa} .
Note here that we use 10000 mapped vectors for construction of a SINDy model.
Using $\alpha=1$, the reproduced trajectory is in excellent agreement with the original one.
However, there are many terms in the ordinary differential equations and some coefficients are too large {although the oscillation are presented with a shallow range ($-1, 1$)}.
The result with the TLSA here is likely because we do not use the data processing, which causes the large regression coefficients and non-sparse results.
On the other hand, the reproduced trajectory with $\alpha=10$ converges to a point around $(r_1,r_2)=(0.6,0)$, although the model becomes parsimonious.
Since the predicted latent vector converges after $t\approx 4$ as shown in the time history of figure \ref{fig_per_tlsa}, the temporal evolution of the flow field by the CNN decoder is also frozen {as shown in figure \ref{fig_per_tlsa_field}}.
Other observation here is that there remains no terms in the ordinary differential equations with high threshold, i.e., $\alpha \geq 50$.
\begin{figure}
  \centerline{\includegraphics[clip,width=0.7\linewidth]{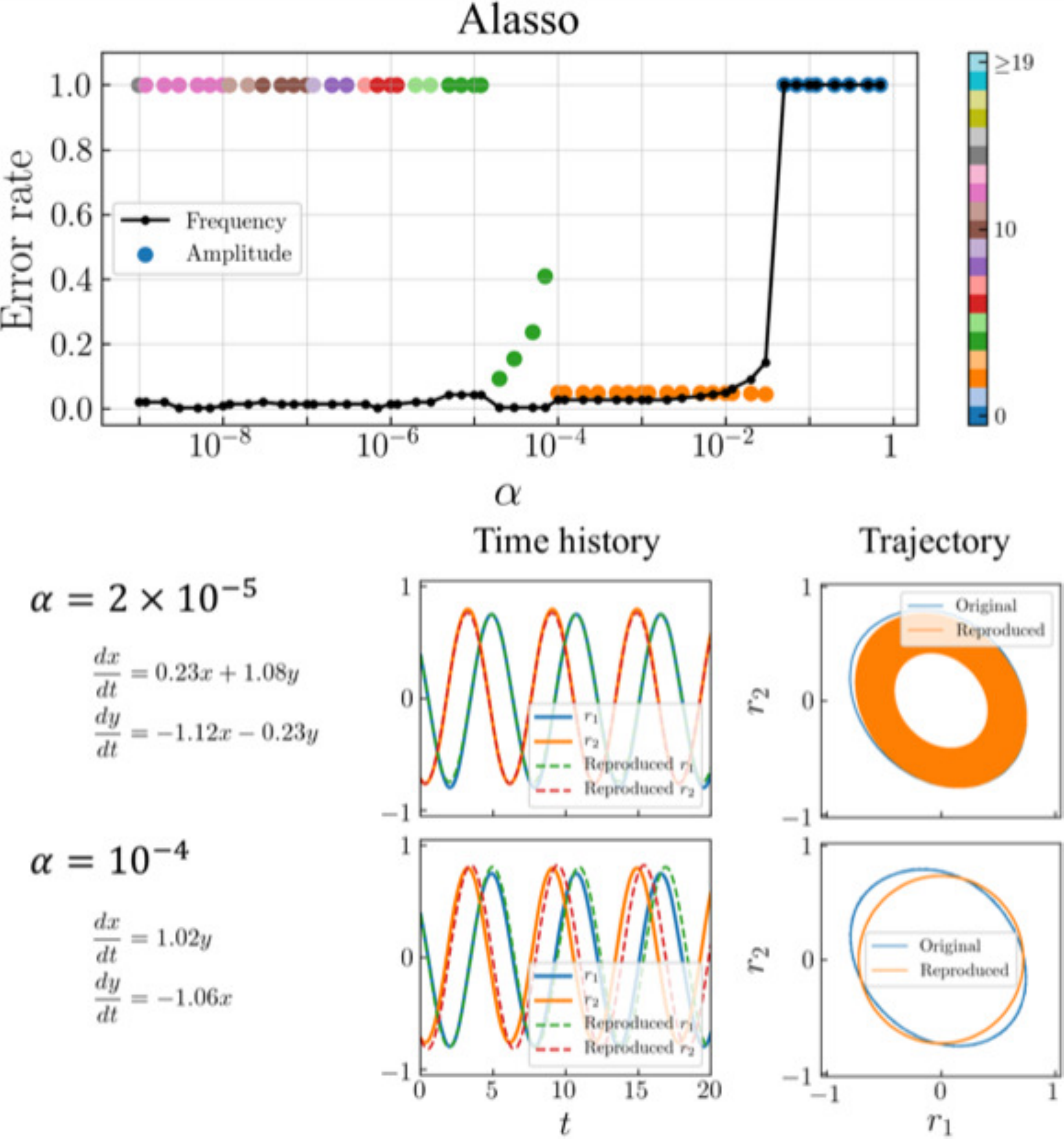}}
  \caption{Results with Alasso of a periodic cylinder example.
  Parameter search results, examples of obtained equations, and reproduced trajectories with $\alpha=\SI{2e-5}{}$ and $\alpha=\SI{1e-4}{}$ are shown. Color of amplitude plot indicates the total number of terms. In the ordinary differential equations, the latent vector components $(r_1, r_2)$ are represented by $(x,y)$ for the distinctness.
  }
  \label{fig_per_alasso}
\end{figure}

\begin{figure}
  \centerline{\includegraphics[clip,width=0.6\linewidth]{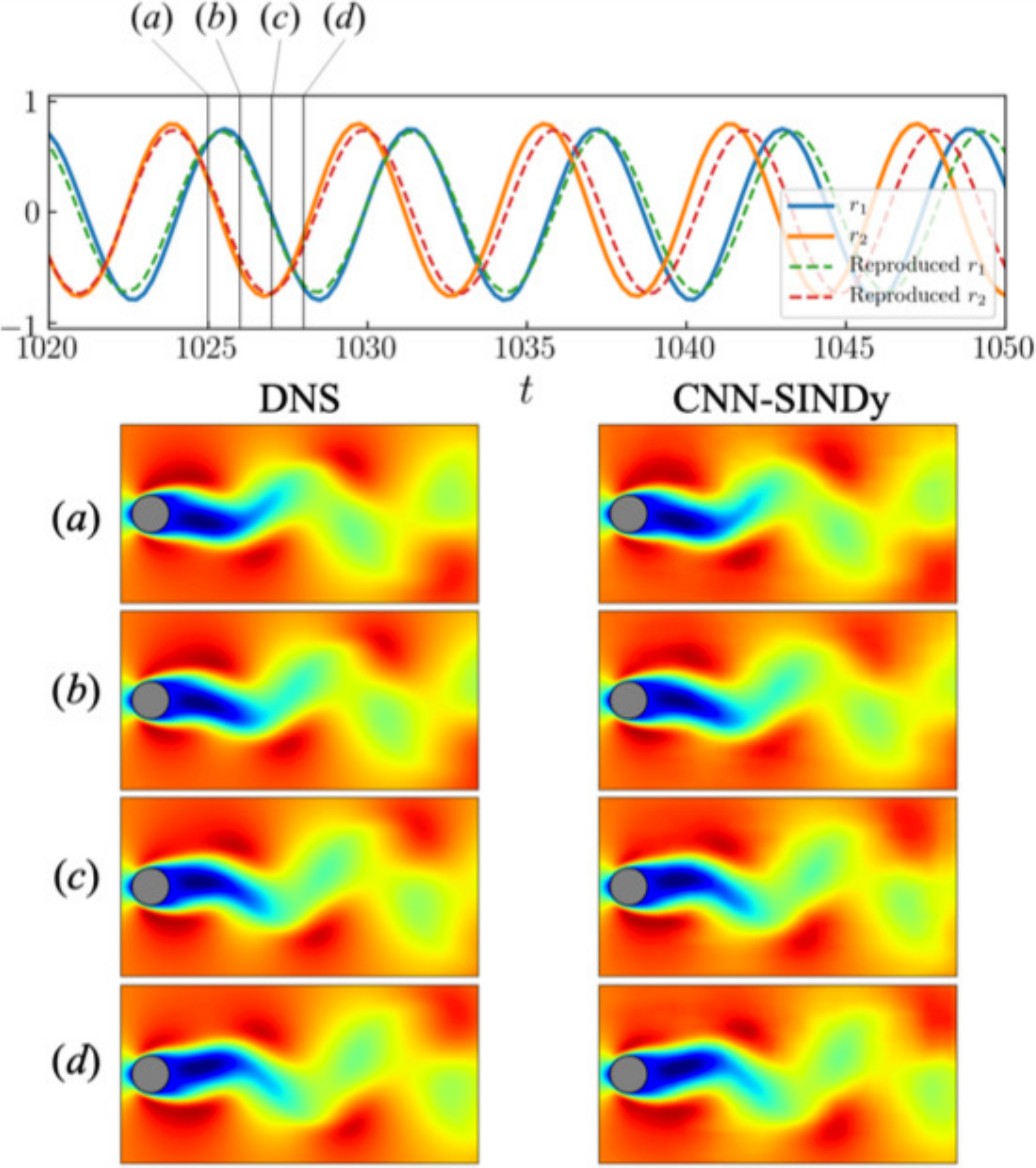}}
  \caption{Time history of the latent vector and temporal evolution of the wake of DNS and the reproduced field at $(a)$ $t=1025$, $(b)$ $t=1026$, $(c)$ $t=1027$ and $(d)$ $t=1028$.}
  \label{fig_per_alasso_field}
\end{figure}

With Alasso as the regression method, we can find the candidate models with low errors at $\alpha=\SI{2e-5}{}$ and $\SI{1e-4}{}$ as shown in figure \ref{fig_per_alasso}.
At $\alpha=\SI{2e-5}{}$, the ordinary differential equations consist of four coefficients in total and the reproduced trajectory gradually converges as shown.
On the other hand, by choosing the appropriate sparsity constant, i.e., $\alpha = \SI{1e-4}{}$, the circle-like {oscillating} trajectory, which is similar to the reference, can be represented with only two terms.
We then check the reproduced flow field by the combination of the predicted latent vector by SINDy and CNN decoder, as shown in figure \ref{fig_per_alasso_field}.
The temporal evolution of high-dimensional dynamics can be reproduced well using the proposed model, although the period of two fields are slightly different.

\begin{figure}
  \centerline{\includegraphics[clip,width=0.7\linewidth]{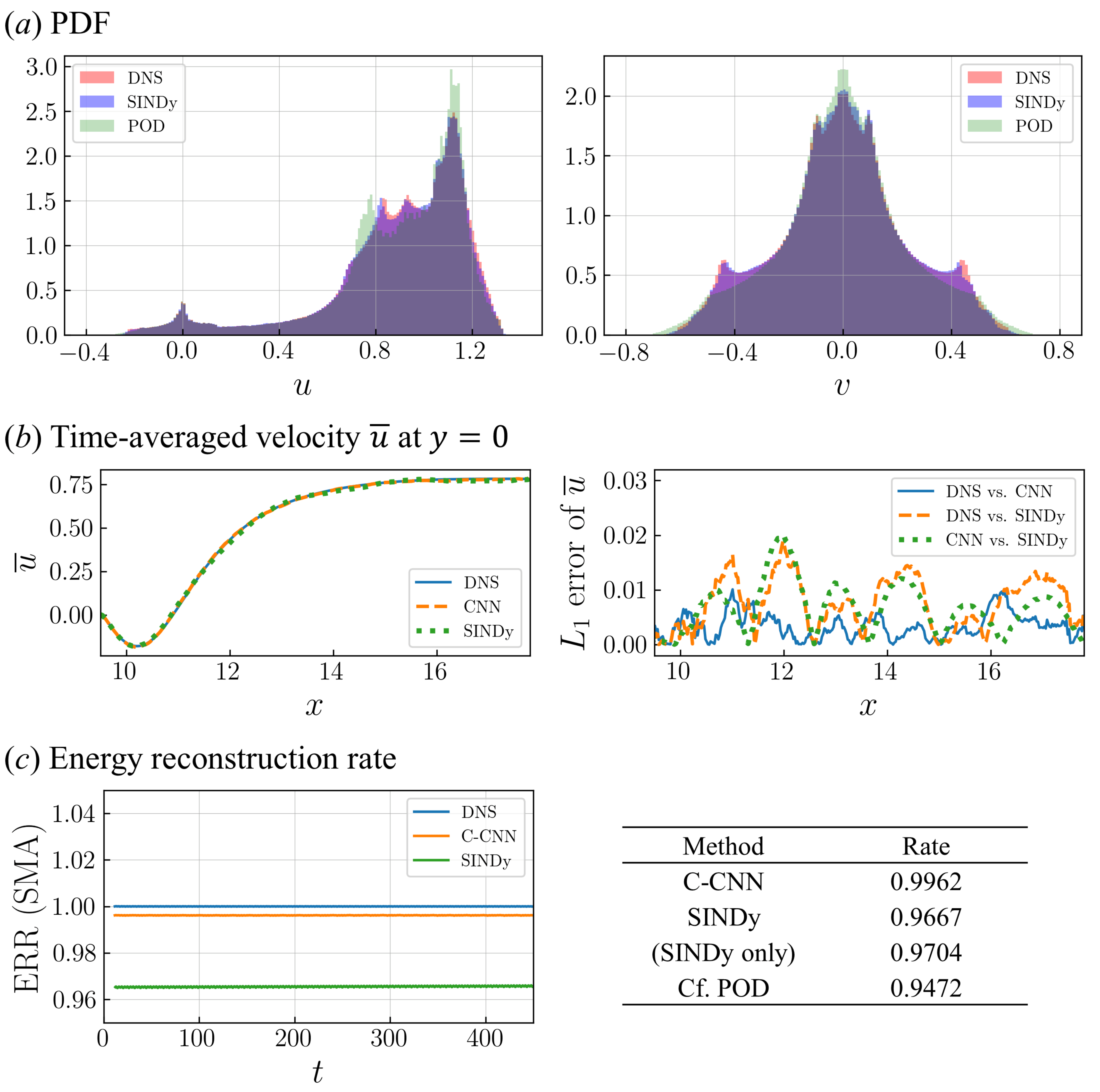}}
  \caption{Evaluation of reproduced flow field with a periodic shedding. $(a)$ Probability density function, $(b)$ mean streamwise velocity at $y=0$, and $(c)$ the energy reconstruction rate (ERR). Simple moving average (SMA) of ERR is shown here for the clearness.}
  \label{fig_per_alasso_eval}
\end{figure}

The quantitative analysis with the probability density function (PDF), the mean streamwise velocity at $y=0$, and the time history of energy reconstruction rate is summarized in figure \ref{fig_per_alasso_eval}.
The energy reconstruction rate ${\cal R}$ is defined as
\begin{align}
    {\cal R}= \dfrac{ \int_S (u^{\prime 2}_{\rm Rep}+{v^{\prime}}^2_{\rm Rep})dS }{ \int_S (u^{\prime 2}_{\rm DNS}+v^{\prime 2}_{\rm DNS}) dS },
\end{align}
where $(u'_{\rm Rep},v'_{\rm Rep})$ and $(u'_{\rm DNS}, v'_{\rm DNS}$) denote the fluctuation components of the reproduced velocity and the original velocity, respectively.
For the PDF, the distribution of CNN-SINDy model is in great agreement with the reference DNS data.
Since the SINDy model of the present case integrates the latent vector via the CNN-AE which can map a high-dimensional flow field into low-dimensional space efficiently thanks to nonlinear activation function \citep{MFF2020}, the SINDy outperforms POD with 2 modes as long as the appropriate waveforms can be obtained. 
In other words, the obtained equation can be integrated stably.
The success of the CNN-SINDy based modeling can be also seen with time-averaged velocity and energy containing rate in figure \ref{fig_per_alasso_eval}.

\subsection{Example 2: transient wake of cylinder flow}
\label{tcw}

\begin{figure}
  \centerline{\includegraphics[clip,width=0.8\linewidth]{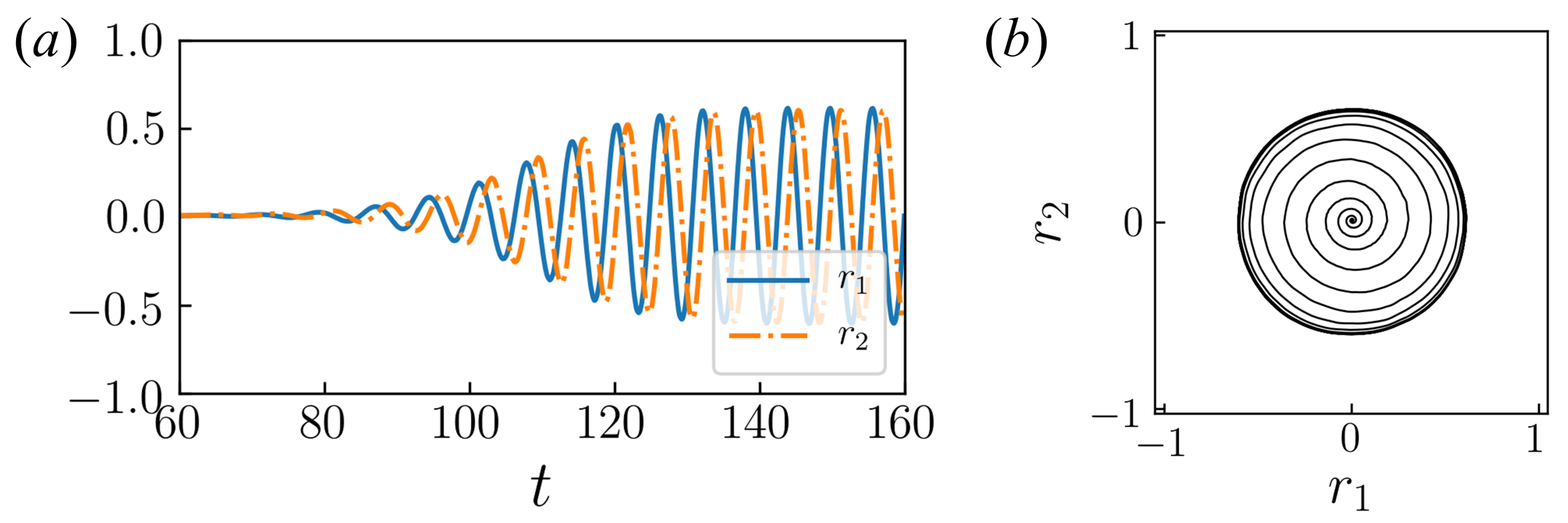}}
  \caption{Latent vector dynamics of the transient process. ($a$) time history and ($b$) trajectory.}
  \label{fig_trans}
\end{figure}

As the second example for the combination of CNN and SINDy, we consider the transient process of a flow around a circular cylinder at $\Rey_D=100$. 
For taking into account the transient process, the streamwise length of computational domain and that of flow field data are extended to $L_x=51.2$ and $8.2 \leq x \leq 37$, i.e., $N_x^*=1152$, same setup with~\citet{MFF2020}.
The time history and the trajectory in the latent space are shown in figure \ref{fig_trans}.
Since the trajectory looks like a circle as shown in figure \ref{fig_trans}$(b)$, we use the residual sum of error of $r_1^2+r_2^2$ between the original value and the reproduced value as the evaluation index in this example.

For SINDy, we use the data with $t=[60,160]$.
The library matrix contains up to 5th order terms. 
Although equations, which can provide the correct temporal evolution of the latent vector, can be obtained with Alasso by including higher order terms, the model here is, of course, not sparse and difficult to interpret.

\begin{figure}
  \centerline{\includegraphics[clip,width=1.0\linewidth]{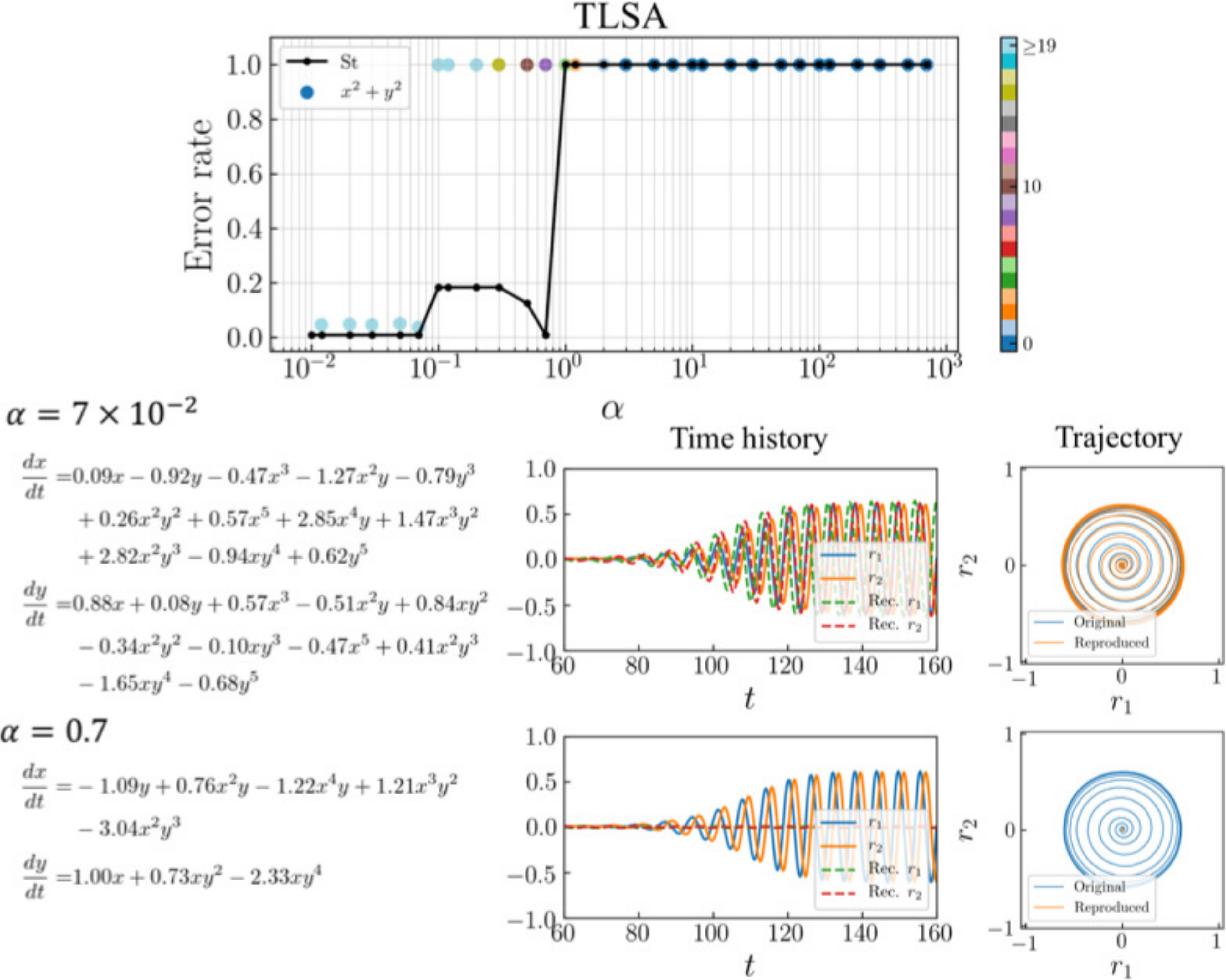}}
  \caption{Results with TLSA of the transient example.
  Parameter search results, examples of obtained equations, and reproduced trajectories with $\alpha=\SI{7e-2}{}$ and $\alpha=\SI{0.7}{}$ are shown. Color of amplitude plot indicates the total number of terms. In the ordinary differential equations, the latent vector components $(r_1, r_2)$ are represented by $(x,y)$ for the distinctness.
  }
  \label{fig_trans_tlsa}
\end{figure}

The parameter search results with TLSA for transient wake are summarized in figure~\ref{fig_trans_tlsa}.
Since the trajectory here is simple as shown, we can obtain the governing equations, which provide the reasonable agreement with the reference, at $\alpha=7\times 10^{-2}$.
With a higher sparsity constant, i.e., $\alpha=0.7$, the equation provides a temporal evolution of the latent vector with almost no oscillation.

\begin{figure}
  \centerline{\includegraphics[clip,width=1.0\linewidth]{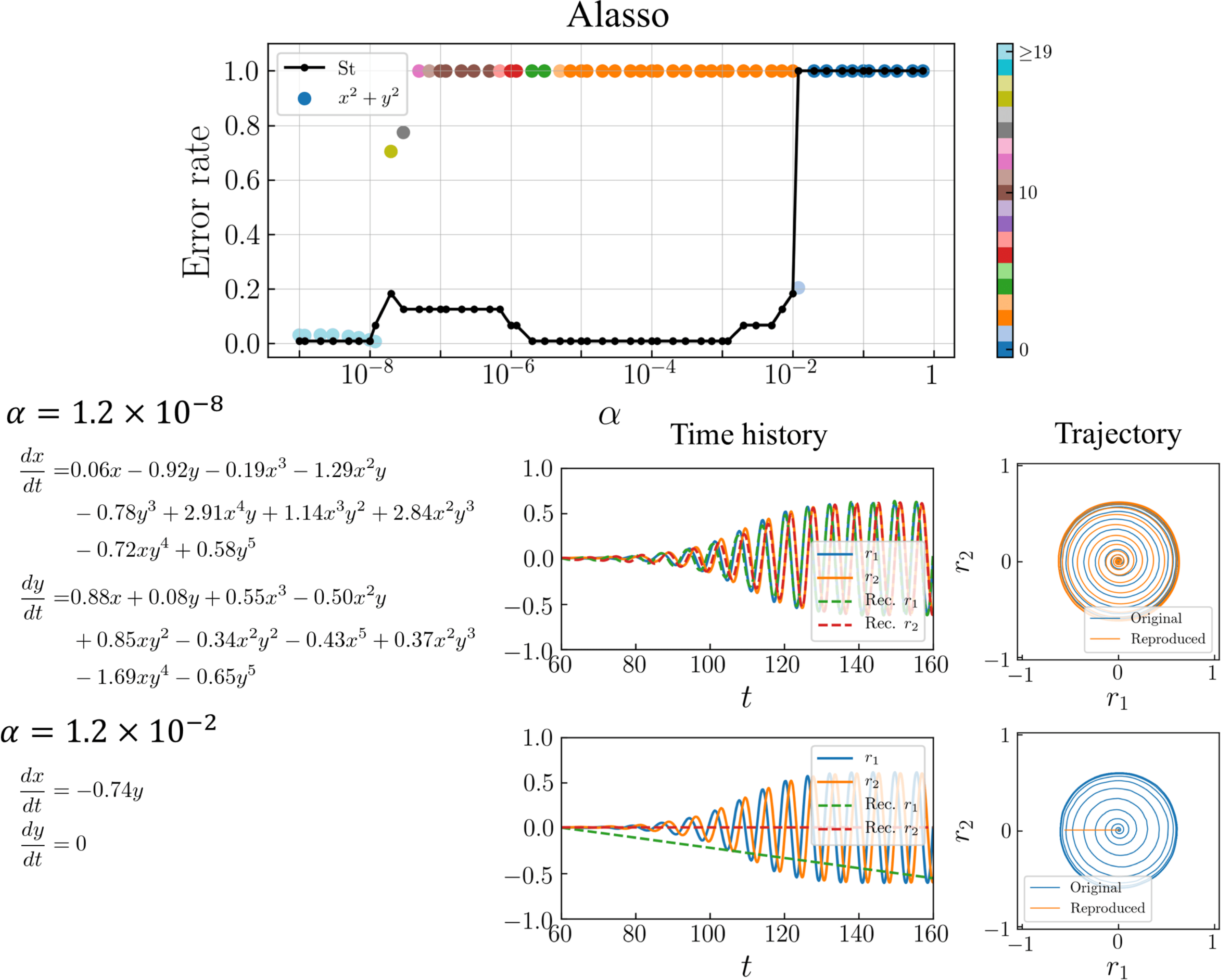}}
  \caption{Results with Alasso of the transient example.
  Parameter search results, examples of obtained equations, and reproduced trajectories with $\alpha=\SI{1.2e-8}{}$ and $\alpha=\SI{1.2e-2}{}$ are shown. Color of amplitude plot indicates the total number of terms. In the ordinary differential equations, the latent vector components $(r_1, r_2)$ are represented by $(x,y)$ for the distinctness.
  }
  \label{fig_trans_alasso}
\end{figure}

Alasso is also considered as shown in figure~\ref{fig_trans_alasso}.
Although the number of terms is larger than that in the periodic shedding example, the trajectory can be reproduced successfully at $\alpha=\SI{1.2e-8}{}$.
With a higher sparsity constant, e.g., $\alpha=\SI{1.2e-2}{}$, the developing oscillation is not reproduced due to a lack of number of terms.
Noteworthy here is that the error for frequency is almost negligible in the range of $5\times 10^{-6}\leq \alpha \leq 10^{-3}$ in spite of the high error rate regarding $x^2+y^2$.
This is because the slight oscillation occurs with the similar frequency to the solution.
Although the identified equation here is more complex than the well-known transient system with a two-dimensional paraboloid manifold~\citep{Loiseau2018,LB2018,sipp2007global}, it is likely caused by the complexity of captured information inside the AE modes compared to the POD modes~\citep{MFF2020}.
This observation enables us to notice the trade-off relationship between the used information associated with the number of modes and the sparsity of equation.
Hence, taking a constraint for nonlinear AE modes to be handy, e.g., orthogonal, may be helpful to identify more sparse dynamics~\citep{CLKB2019,jayaraman2020data,ladjal2019pca}.

\begin{figure}
  \centerline{\includegraphics[clip,width=1\linewidth]{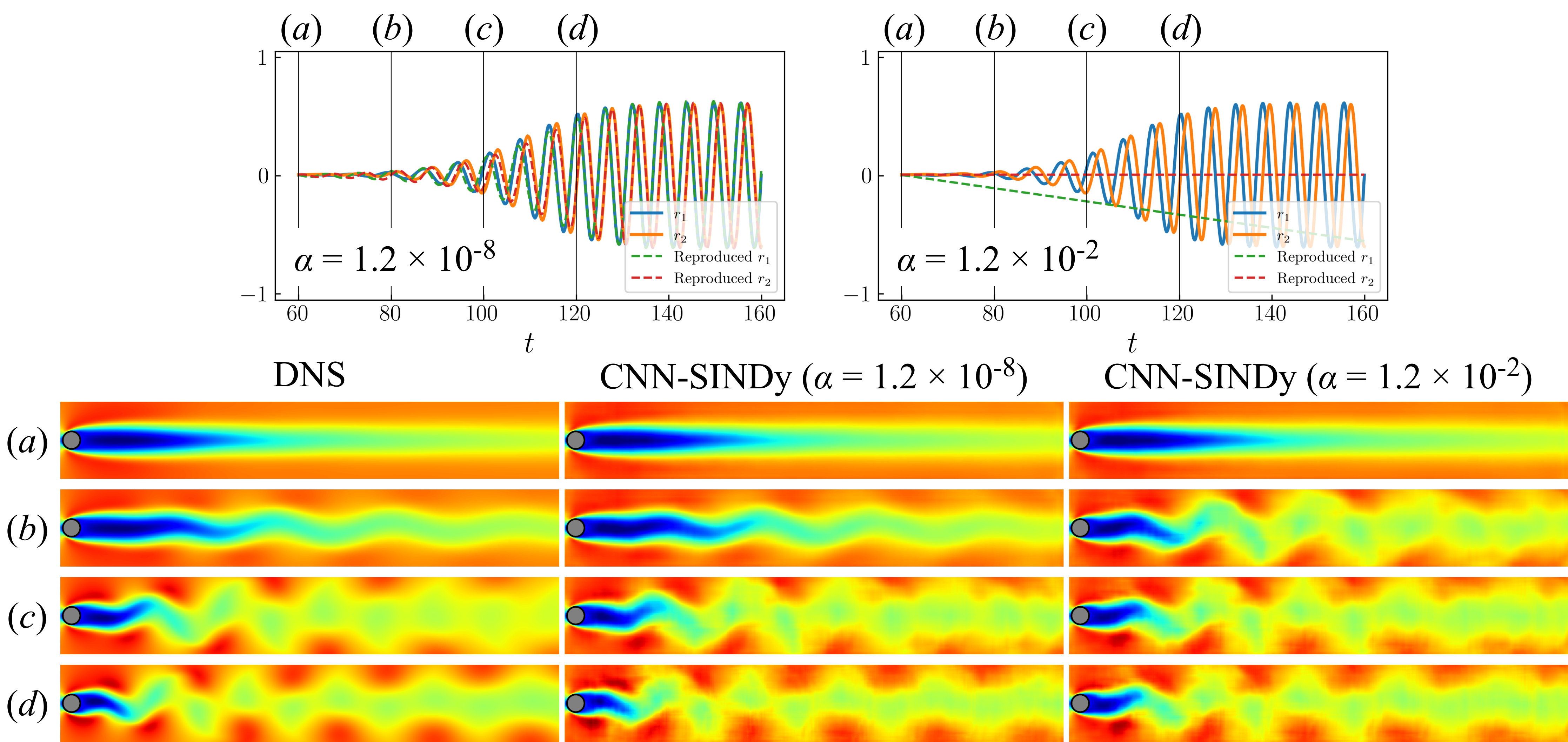}}
  \caption{
  Time history of latent vector and the temporal evolution of the wake of DNS and the reproduced field at $(a)$ $t=60$, $(b)$ $t=80$, $(c)$ $t=100$ and $(d)$ $t=120$.
  }
  \label{fig_trans_alasso_field}
\end{figure}

The streamwise velocity fields reproduced by the CNN-AE and the integrated latent variables with the Alasso-based SINDy are presented in figure~\ref{fig_trans_alasso_field}.
The present two sparsity constants here correspond to the results in figure~\ref{fig_trans_alasso}.
With $\alpha=1.2\times 10^{-8}$, the flow field can successfully be reproduced thanks to nonlinear terms in the identified equations by SINDy.
In contrast, it is tough to capture the transient behavior correctly by the CNN-SINDy model with $\alpha=1.2\times 10^{-2}$, which can explicitly be found by the comparison among $(b)$ in figure~\ref{fig_trans_alasso_field}.  
This corresponds to the observation that the sparse model cannot reproduce the trajectory well in figure~\ref{fig_trans_alasso}.

\subsection{Example 3: two-parallel cylinders wake}
\label{sec:tpcw}

To demonstrate the applicability of the present CNN-SINDy based reduced-order modeling to more complex wake flows,
let us here consider a flow around two-parallel cylinders whose radii are different with each other~\citep{MFZF2020,morimoto2101convolutional}, as presented in figure~\ref{fig1}.
Unlike the wake dynamics of two identical circular cylinders as described in~\citet{kang2003characteristics}, the coupled wakes behind two side-by-side uneven cylinders exhibits more complex vortex interactions, due to the mismatch in their individual shedding frequencies.
Training flow snapshots are prepared with a direct numerical simulation with the open-source CFD toolbox OpenFOAM~\citep{weller1998tensorial}, using second-order discretization schemes in both time and space.
We arrange the two circular cylinders with a size ratio of $r$ and a gap of $gD$, where $g$ is the gap ratio and $D$ is the diameter of the lower cylinder.
The Reynolds number is fixed at $Re_D=100$. 
The two cylinders are placed $20D$ downstream of the inlet where a uniform flow with velocity $U_{\infty}$ is prescribed, and $40D$ upstream of the outlet with zero pressure. 
The side boundaries are specified as slip and are $40D$ apart. 
The time stpdf for the DNS and the snapshot sampling are respectively 0.01 and 0.1.
A notable feature of wakes associated with the present two-cylinder position is complex wake behavior caused by varying the size ratio $r$ and the gap ratio $g$~\citep{MFZF2020,morimoto2101convolutional}.
In the present study, the wake with the combination of $\{r,g\}=\{1.15,2.0\}$, which is a quasi-periodic in time, is considered for the demonstration.
For the training of both the CNN-AE and the SINDy, we use 2000 snapshots.
The vorticity field $\omega$ is used as a target attribute in this example.

\begin{figure}
  \centerline{\includegraphics[width=0.9\linewidth]{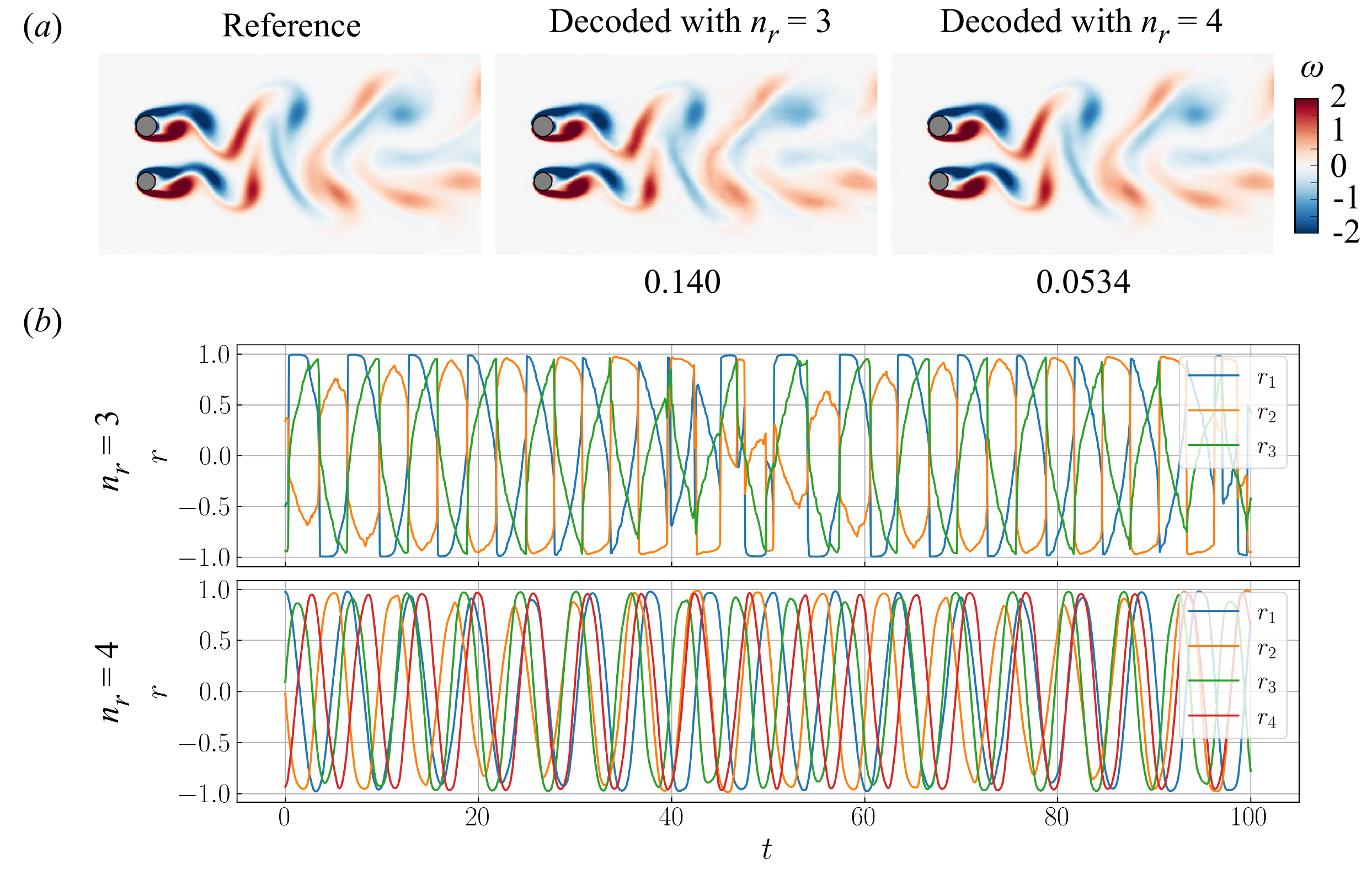}}
  \caption{
  Autoencoder-based low-dimensionalization for the wake of the two-parallel cylinders. $(a)$ Comparison of the reference DNS, the decoded field with $n_r=3$, and the decoded field with $n_r=4$. The values underneath the contours indicate the $L_2$ error norm. $(b)$ Time series of the latent variables with $n_r=3$ and 4.
  }
  \label{fig_AE_2cy}
\end{figure}

Prior to the use of SINDy, let us determine the size of latent variables $n_r$ of this example, as summarized in figure~\ref{fig_AE_2cy}.
We here construct AEs with two $n_r$ cases of 3 and 4.
As presented in figure~\ref{fig_AE_2cy}$(a)$, the recovered flow fields through the AE-based low-dimensionalization are in reasonable agreement with the reference DNS snapshots.
Quantitatively speaking, the $L_2$ error norms between the reference and the decoded fields are approximately 14\% and 5\% for each case.
Although both models achieve the smooth spatial reconstruction, we can find the significant difference by focusing on their temporal behaviors in figure~\ref{fig_AE_2cy}$(b)$.
While the time trace with $n_r=4$ shows smooth variations for all four variables, the curves with $n_r=3$ exhibit a shaky behavior despite the quasi-periodic nature of the flow.
This is caused by the higher error level of the $n_r=3$ case.
Furthermore, it is also known that a high-error level due to an overcompression via AE is highly influential for a temporal prediction of them since there should be {\it exposure bias}~\citep{endo2018multi} occurred by a temporal integration~\citep{HFMF2019,HFMF2020b,nakamura2020extension}.
In what follows, we use the AE with $n_r=4$ for the SINDy-based reduced-order modeling.

\begin{figure}
  \centerline{\includegraphics[width=0.5\linewidth]{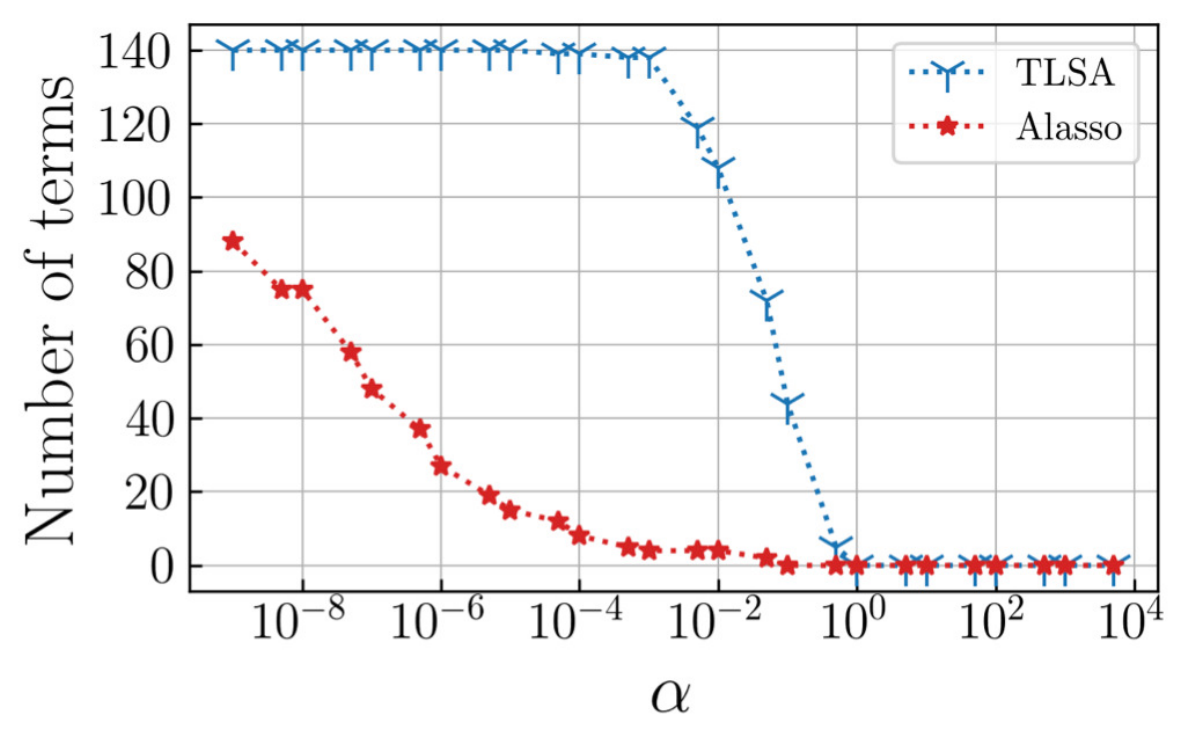}}
  \caption{
  The relationship between the sparsity constant $\alpha$ and the number of terms in identified equations via SINDys with TLSA and Alasso for the two-parallel cylinder wake example.
  }
  \label{fig_2cy_term_SINDy}
\end{figure}

Let us perform SINDy with TLSA and Alasso for the AE latent variables with $n_r=4$.
For the construction of a library matrix of this example, we include up to third-order terms in terms of four latent variables.
The relationship between the sparsity constant $\alpha$ and the number of terms in identified equations is examined in figure~\ref{fig_2cy_term_SINDy}.
Similarly to the previous two examples with a single cylinder, the trends for TLSA and Alasso are different with each other.
Note that this map can only be referred to check the sparsity trend of the identified equation since there is no model equation.
To validate the fidelity of the equations, we need to integrate identified equations and decode a flow field using the decoder part of AE.

\begin{figure}
  \centerline{\includegraphics[width=0.99\linewidth]{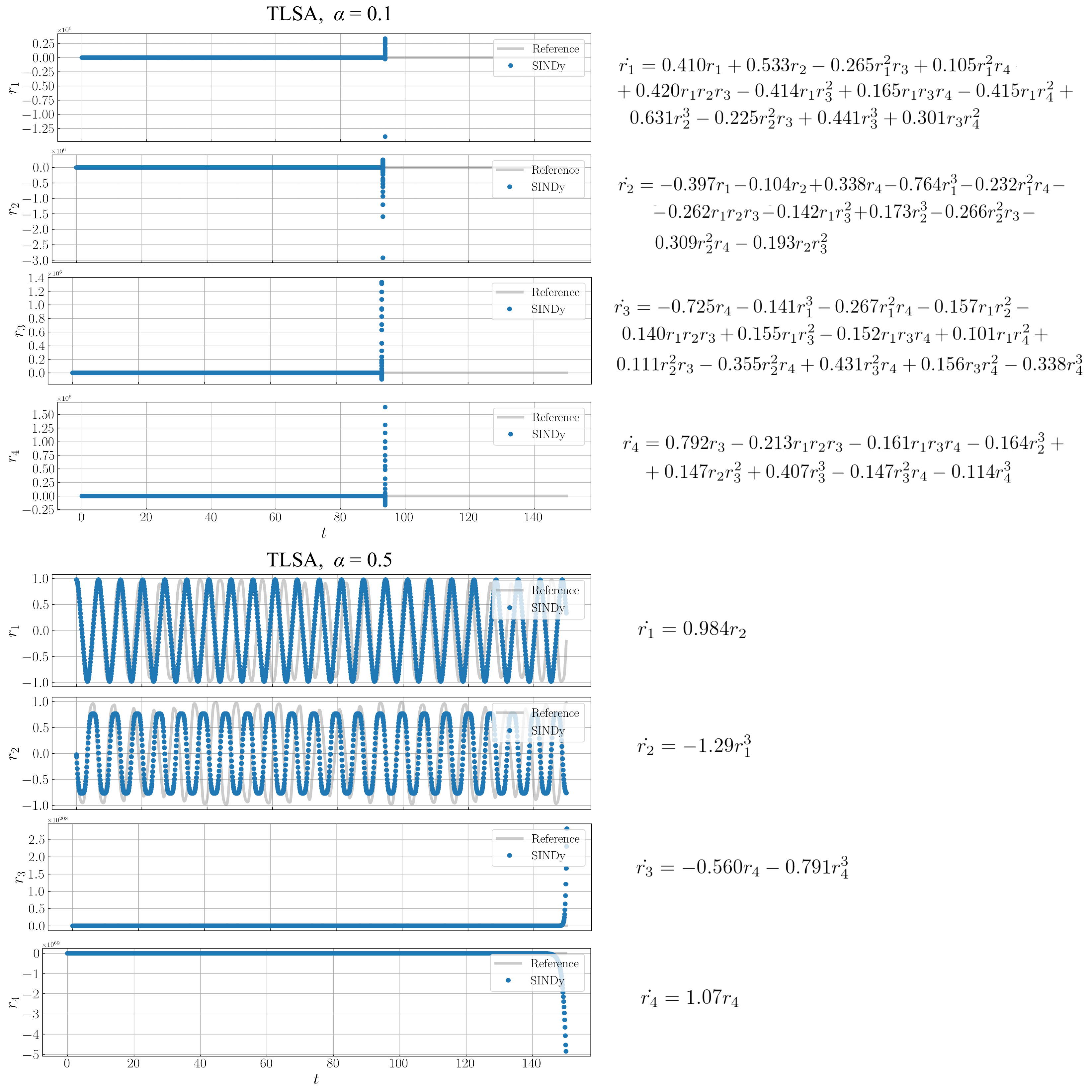}}
  \caption{
  Integration of the identified equations via the TLSA-based SINDy for the two-parallel cylinder wake example. The cases with $\alpha=0.1$ and 0.5 are shown.}
  \label{fig_2cy_TLSA}
\end{figure}

The identified equations via the TLSA-based SINDy are investigated in figure~\ref{fig_2cy_TLSA}.
As examples, the cases with $\alpha=0.1$ and 0.5 are shown.
The case with $\alpha=1$ which contains nonlinear terms shows a diverging
behavior at $t\approx 90$.
The sparse equation obtained with $\alpha=0.5$ can provide stable solutions for $r_1$ and $r_2$; however, its sparseness causes the unstable integration for $r_3$ and $r_4$.
Note in passing that the TLSA-based modeling cannot identify the stable solution although we have carefully examined the influence on the sparsity factors for other cases not shown here.

\begin{figure}
  \centerline{\includegraphics[width=0.99\linewidth]{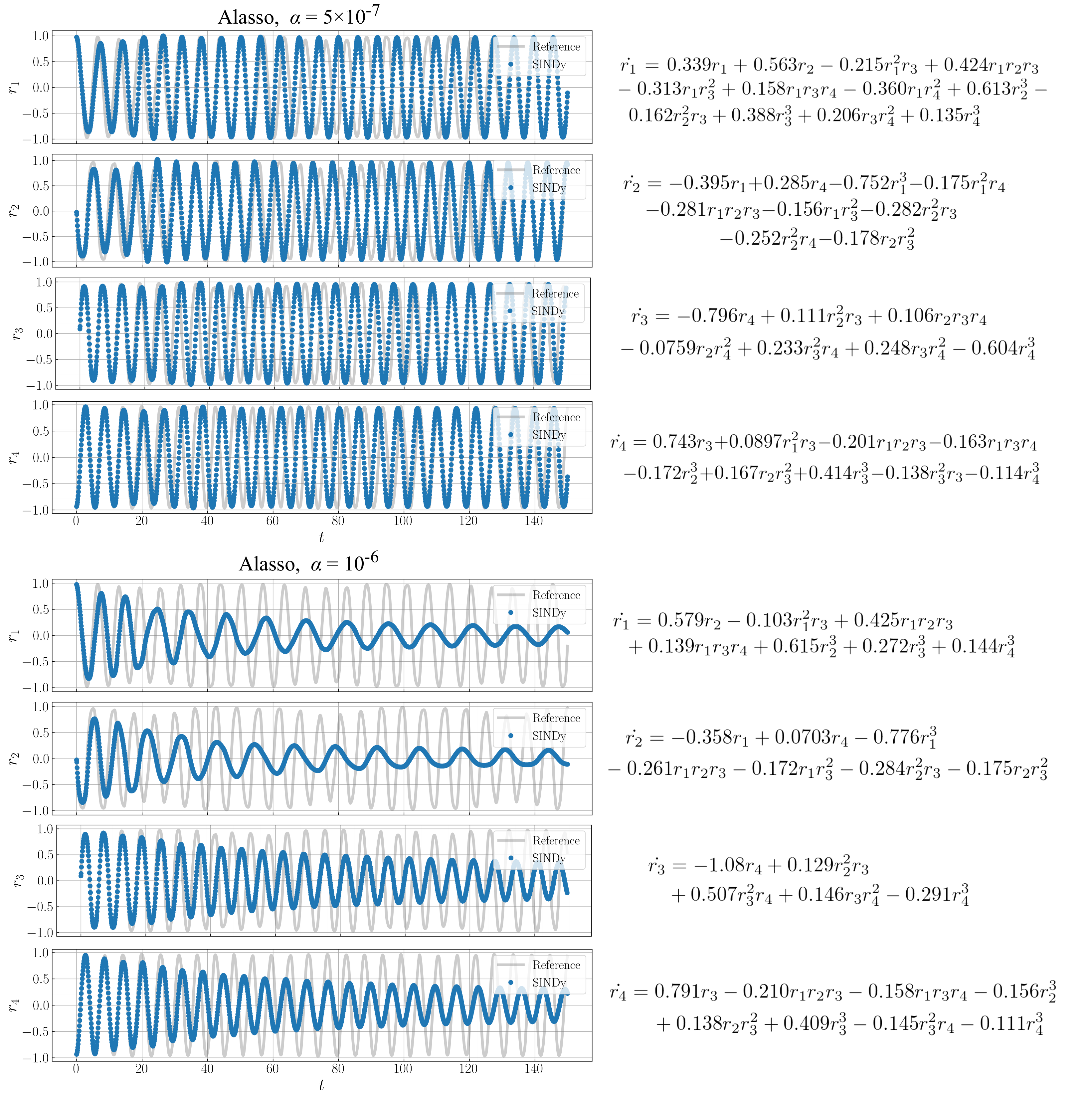}}
  \caption{
  Integration of the identified equations via the Alasso-based SINDy for the two-parallel cylinder wake example. The cases with $\alpha=5\times 10^{-7}$ and $10^{-6}$ are shown.}
  \label{fig_2cy_Alasso}
\end{figure}

We then use Alasso as a regression function of SINDy, as summarized in figure~\ref{fig_2cy_Alasso}.
The cases with $\alpha=5\times 10^{-7}$ and $10^{-6}$ are shown.
The identified equation with $\alpha=5\times 10^{-7}$ can successfully be integrated and its temporal behavior is in good agreement with the reference curve provided by the AE-based latent variables.
However, what is notable here is that the equation with $\alpha=10^{-6}$ shows the decaying behavior with increasing the integration window despite that the equation form is very similar to the case with $\alpha=5\times 10^{-7}$.
Some nonlinear terms are eliminated due to the slight increase in the sparsity constant.
These results indicate the importance of the careful choice for both regression functions and 
the sparsity constants associated with them.

\begin{figure}
  \centerline{\includegraphics[width=0.85\linewidth]{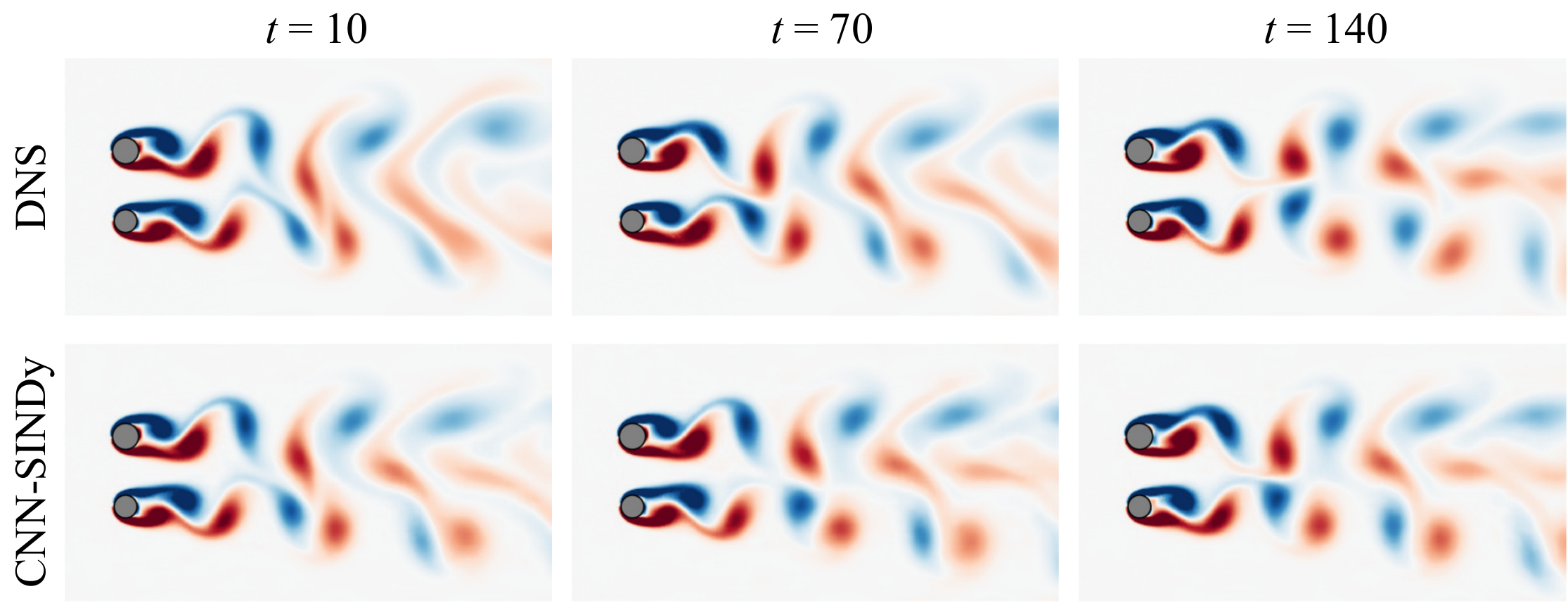}}
  \caption{
  Reproduced fields with the CNN-SINDy based reduced-order modeling of the two-parallel cylinders example. The case of Alasso with $\alpha=5\times 10^{-7}$ is used for SINDy. DNS flow fields are also shown for the comparison.}
  \label{fig_2cy_ROM}
\end{figure}

Based on the results above, the integrated variables with the case of Alasso with $\alpha=5\times 10^{-7}$ are fed into the CNN decoder, as presented in figure~\ref{fig_2cy_ROM}.
The reproduced flow fields are in reasonable agreement with the reference DNS data, although there is a slight offset due to numerical integration.
Hence, the present reduced-order model is able to achieve a reasonable wake reconstruction by following only the temporal evolution of low-dimensionalized vector and caring the selection of parameters, which is akin to the observation of single cylinder cases.

\subsection{Outlook: nine-equation shear flow model}
\label{sec:nine-eq}

One of the remaining issues of the present CNN-SINDy model is the applicability to flows where a lot of spatial mode are required to reconstruct the flow, e.g., turbulence.
With the current scheme for the CNN-AE, it is tough to compress turbulent flow data while keeping the information of high-dimensional dynamics~\citep{MFF2020}.
We have also recently reported that the toughness for turbulence low-dimensionalization is still remaining even if we use a customized autoencoder, although it can achieve the better mapping than that by conventional AE and POD~\citep{FNF2020}.
In addition, the number of terms in a coefficient matrix must be drastically increased for the turbulent case unless the flow can be expressed with few number of modes.
Hence, our next question here is ``Can we also use SINDy if a turbulent flow can be mapped into few number modes?".
In this section, let us consider a nine-equation turbulent shear flow model between infinite parallel free-slip walls under a sinusoidal body force~\citep{MFE2004} as the preliminary example for the application to turbulence with low number modes.

In the nine-equation model, various statistics, including mean velocity profile, streaks, and vortex structures, can be represented with only nine Fourier modes ${\bm u}_j({\bm x})$. 
Analogous to POD, the flow fields can be mathematically expressed with superposition of temporal coefficient and mode such that
\begin{eqnarray}
    {\bm u}({\bm x},t) = \sum_{j=1}^9 a_j(t){\bm u}_j({\bm x}).
\end{eqnarray}
Here, nine ordinary differential equations for the nine mode coefficients are as follows:
\begin{eqnarray}
\dfrac{{\rm d}a_1}{{\rm d}t}&=&\dfrac{\mu^2}{Re}-\dfrac{\mu^2}{Re}a_1-\sqrt{\dfrac{3}{2}}\dfrac{\mu \gamma}{\kappa_{\zeta\mu\gamma}}a_6a_8+\sqrt{\dfrac{3}{2}}\dfrac{\mu \gamma}{\kappa_{\mu\gamma}}a_2a_3
\end{eqnarray}
\vspace{-5mm}
\begin{eqnarray}
\dfrac{{\rm d}a_2}{{\rm d}t}&=&-Re^{-1}\biggl(\dfrac{4\mu^2}{3}+\gamma\biggr)a_2+\dfrac{5\sqrt{2}}{3\sqrt{3}}\dfrac{\gamma^2}{\kappa_{\zeta\gamma}}a_4a_6-\dfrac{\gamma^2}{\sqrt{6}\kappa_{\zeta\gamma}}a_5a_7-\dfrac{\zeta\mu\gamma}{\sqrt{6}\kappa_{\zeta\gamma}\kappa_{\zeta\mu\gamma}}a_5a_8\nonumber\\
&-&\sqrt{\dfrac{3}{2}}\dfrac{\mu\gamma}{\kappa_{\mu\gamma}}a_1a_3-\sqrt{\dfrac{3}{2}}\dfrac{\mu\gamma}{\kappa_{\mu\gamma}}a_3a_9
\end{eqnarray}
\vspace{-5mm}
\begin{eqnarray}
\dfrac{{\rm d}a_3}{{\rm d}t}&=&-\dfrac{\mu^2+\gamma^2}{Re}a_3+\dfrac{2}{\sqrt{6}}\dfrac{\zeta\mu\gamma}{\kappa_{\zeta\gamma}\kappa_{\mu\gamma}}(a_4a_7+a_5a_6)\nonumber\\
&~&~~~~~~~~~~~~~~~~~~~~~~~~~~~~~~~~+\dfrac{\mu^2(3\zeta^2+\gamma^2)-3\gamma^2(\zeta^2+\gamma^2)}{\sqrt{6}\kappa_{\zeta\gamma}\kappa_{\mu\gamma}\kappa_{\zeta\mu\gamma}}a_4a_8
\end{eqnarray}
\vspace{-5mm}
\begin{eqnarray}
\dfrac{{\rm d}a_4}{{\rm d}t}&=&-\dfrac{3\zeta^2+4\mu^2}{3Re}a_4-\dfrac{\zeta}{\sqrt{6}}a_1a_5-\dfrac{10}{3\sqrt{6}}\dfrac{\zeta^2}{\kappa_{\zeta\gamma}}a_2a_6\nonumber\\
&~&~~~~~~~~~~~~-\sqrt{\dfrac{3}{2}}\dfrac{\zeta\mu\gamma}{\kappa_{\zeta\gamma}\kappa_{\mu\gamma}}a_3a_7-\sqrt{\dfrac{3}{2}}\dfrac{\zeta^2\mu^2}{\kappa_{\zeta\gamma}\kappa_{\mu\gamma}\kappa_{\zeta\mu\gamma}}a_3a_8-\dfrac{\zeta}{\sqrt{6}}a_5a_9
\end{eqnarray}
\begin{eqnarray}
\dfrac{{\rm d}a_5}{{\rm d}t}&=&-\dfrac{\zeta^2+\mu^2}{Re}a_5+\dfrac{\zeta}{\sqrt{6}}a_1a_4+\dfrac{\zeta^2}{\sqrt{6}\kappa_{\zeta\gamma}}a_2a_7\nonumber\\
&~&~~~~~~~~~~~~~~-\dfrac{\zeta\mu\gamma}{\sqrt{6}\kappa_{\zeta\gamma}\kappa_{\zeta\mu\gamma}}a_2a_8+\dfrac{\zeta}{\sqrt{6}}a_4a_9+\dfrac{2}{\sqrt{6}}\dfrac{\zeta\mu\gamma}{\kappa_{\zeta\gamma}\kappa_{\mu\gamma}}a_3a_6\\
\dfrac{{\rm d}a_6}{{\rm d}t}&=&-\dfrac{3\zeta^2+4\mu^2+3\gamma^2}{3Re}a_6+\dfrac{\zeta}{\sqrt{6}}a_1a_7+\sqrt{\dfrac{3}{2}}\dfrac{\mu\gamma}{\kappa_{\zeta\mu\gamma}}a_1a_8\nonumber\\
&~&+\dfrac{10}{3\sqrt{6}}\dfrac{\zeta^2-\gamma^2}{\kappa_{\zeta\gamma}}a_2a_4-2\sqrt{\dfrac{2}{3}}\dfrac{\zeta\mu\gamma}{\kappa_{\zeta\gamma}\kappa_{\mu\gamma}}a_3a_5+\dfrac{\zeta}{\sqrt{6}}a_7a_9+\sqrt{\dfrac{3}{2}}\dfrac{\mu\gamma}{\kappa_{\zeta\mu\gamma}}a_8a_9
\end{eqnarray}
\begin{eqnarray}
\dfrac{{\rm d}a_7}{{\rm d}t}&=&-\dfrac{\zeta^2+\mu^2+\gamma^2}{Re}a_7-\dfrac{\zeta}{\sqrt{6}}(a_1a_6+a_6a_9)\nonumber\\
&~&~~~~~~~~~~~~~~~~~~~~+\dfrac{1}{\sqrt{6}}\dfrac{\gamma^2-\zeta^2}{\kappa_{\zeta\gamma}}a_2a_5+\dfrac{1}{\sqrt{6}}\dfrac{\zeta\mu\gamma}{\kappa_{\zeta\gamma}\kappa_{\mu\gamma}}a_3a_4\\
\dfrac{{\rm d}a_8}{{\rm d}t}&=&-\dfrac{\zeta^2+\mu^2+\gamma^2}{Re}a_8+\dfrac{2}{\sqrt{6}}\dfrac{\zeta\mu\gamma}{\kappa_{\zeta\gamma}\kappa_{\zeta\mu\gamma}}a_2a_5+\dfrac{\gamma^2(3\zeta^2-\mu^2+3\gamma^2)}{\sqrt{6}\kappa_{\zeta\gamma}\kappa_{\mu\gamma}\kappa_{\zeta\mu\gamma}}a_3a_4\\
\dfrac{{\rm d}a_9}{{\rm d}t}&=&-\dfrac{9\mu^2}{Re}a_9+\sqrt{\dfrac{3}{2}}\dfrac{\mu\gamma}{\kappa_{\mu\gamma}}a_2a_3-\dfrac{\mu\gamma}{\kappa_{\zeta\mu\gamma}}a_6a_8,
\end{eqnarray}
where $\zeta$, $\mu$, and $\gamma$ are constant values, $\kappa_{\zeta\gamma}=\sqrt{\zeta^2+\gamma^2}$, $\kappa_{\mu\gamma}=\sqrt{\mu^2+\gamma^2}$, $\kappa_{\zeta\mu\gamma}=\sqrt{\zeta^2+\mu^2+\gamma^2}$.
These coefficients are multiplied to corresponding Fourier modes which have individual role to reconstruct a flow, e.g., basic profile, streak, and spanwise flows.
We refer enthusiastic readers to \citet{MFE2004} for details.

\begin{figure}
  \centerline{\includegraphics[clip,width=0.95\linewidth]{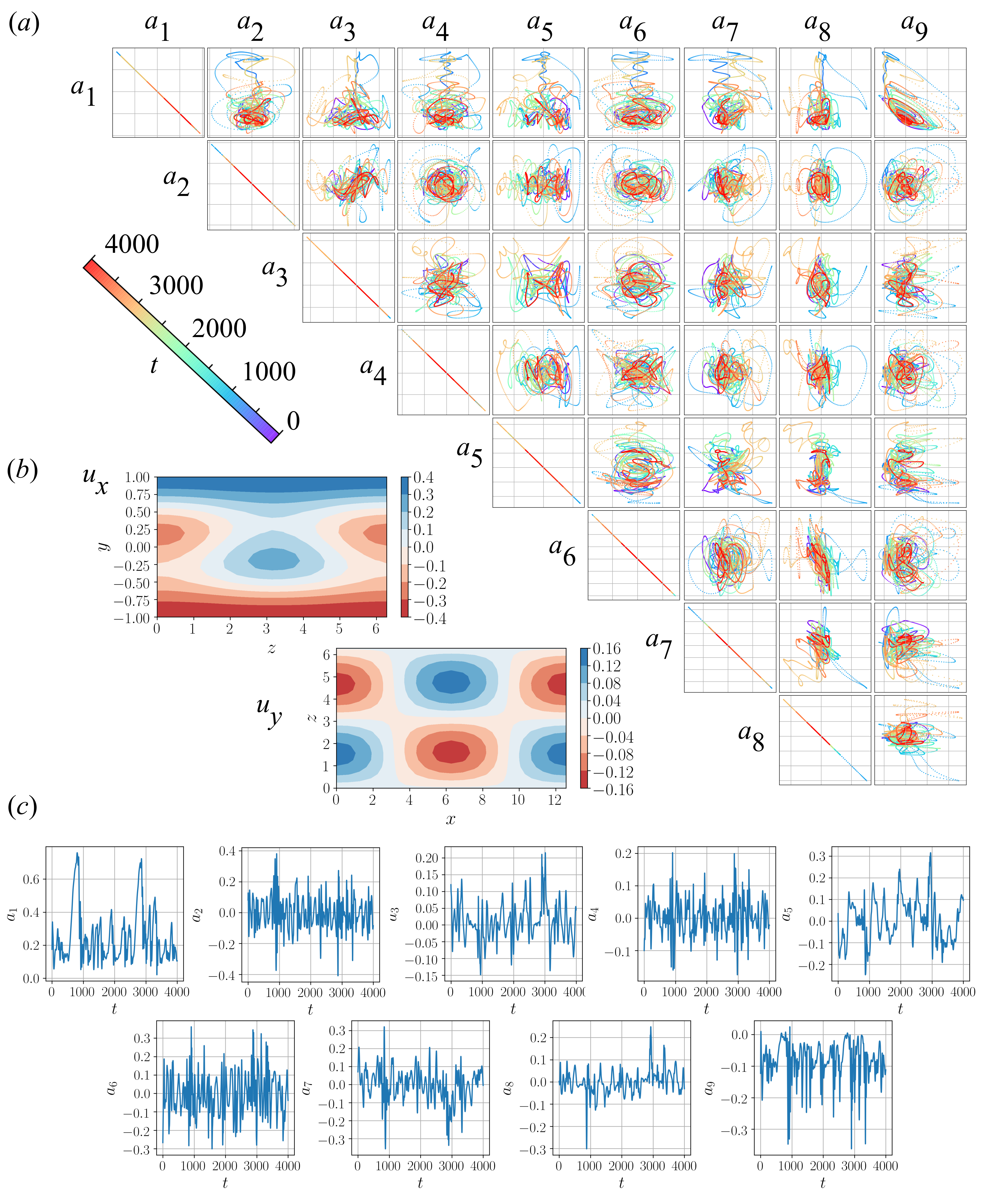}}
  \caption{$(a)$ Pairwise correlations of nine coefficients. $(b)$ Example contours of velocity $u_x$ and velocity $u_y$ at midplane.  $(c)$ Temporal evolution of amplitudes $a_i$.}
  \label{fig_f3}
\end{figure}

In this section, we aim to obtain the coefficient matrix for the simultaneous time differential equations (4.4) to (4.12) using SINDy.
In the following, let us consider the Reynolds number based on the channel half height $\delta$ and laminar velocity $U_0$ at a distance of $\delta/2$ from the top wall set to $Re = 400$.
The initial condition for numerically integration of the equations above is $(a^0_1,a^0_1,a^0_2,a^0_3,a^0_4,a^0_5,a^0_6,a^0_7,a^0_8,a^0_9)=(1, 0.07066, -0.07076, 0, 0, 0, 0, 0)$ with a random small perturbation for $a_4$, which is the same set up as the reference code by \citet{SGASV2019}.
The lengths and the number of grids of the computational domain are set to $(L_x, L_y, L_z)=(4\pi, 2, 2\pi)$ and $(N_x,N_y,N_z)=(21,21,21)$, respectively.
The constant values are set to $(\zeta, \mu, \gamma)=(2\pi/L_x, \pi/2, 2\pi/L_z)$ and the time step is 0.5.
Examples of streamwise-averaged velocity $u_x$ contour and velocity $u_y$ contour at the midplane with the temporal evolution of the amplitudes $a_i$ and the pairwise correlations of the present nine coefficients are shown in figure \ref{fig_f3}.
The chaotic nature of the considered problem can be seen.
For performing SINDy, we use 10000 discretized coefficients as the training data.

\begin{figure}
  \centerline{\includegraphics[clip,width=0.9\linewidth]{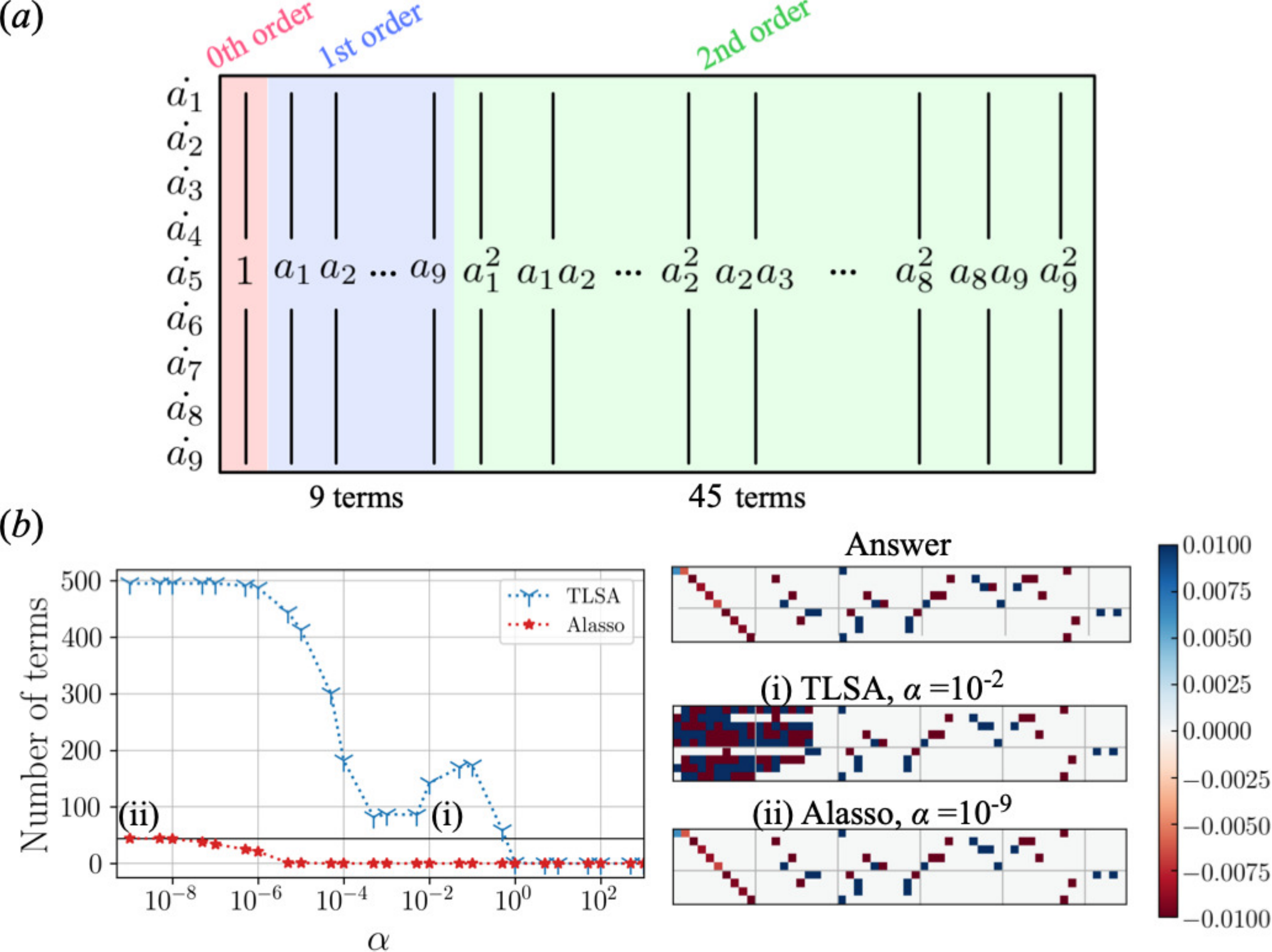}}
  \caption{SINDy for the nine-equation shear flow model. $(a)$ Schematic of coefficient matrix $\beta$. $(b)$ Relationship between the sparsity constant $\alpha$ and the number of terms with the obtained coefficient matrices.}
  \label{fig_f4}
\end{figure}

\begin{figure}
  \centerline{\includegraphics[clip,width=1.0\linewidth]{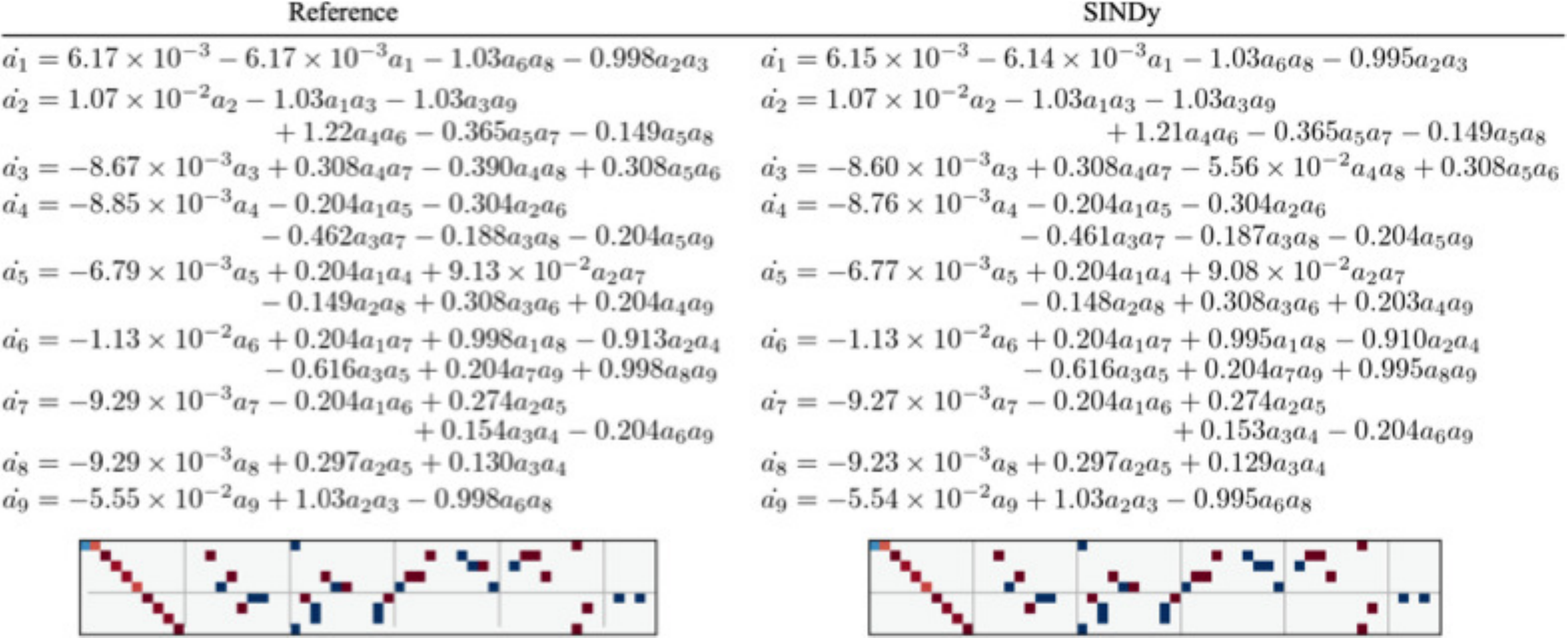}}
  \caption{Comparison of the governing equation for temporal coefficients.}
  \label{fig_f5}
\end{figure}

The SINDy in this section is also performed with TLSA and Alasso following the discussions above.
Since the equations for the temporal coefficients are constructed up to 2nd order terms, the coefficient matrix also includes up to 2nd order terms, as shown in figure~\ref{fig_f4}$(a)$.
The total number of terms considered here is 55.
The results with TLSA and Alasso are summarized in figure~\ref{fig_f4}$(b)$.
The matrices located in the right area correspond to the coefficient matrix in figure~\ref{fig_f4}$(a)$.
Similar to the results above, the model using TLSA has some huge values due to lack of penalty terms.
Especially, the effects by overfitting can be seen for the low-order portion.
On the other hand, the remarkable ability of the SINDy can be seen with the Alasso.
By giving the appropriate sparsity constant $\alpha$, the governing equations can be represented successfully.
The details of each magnitude of coefficients are shown in figure~\ref{fig_f5}.
It is striking that the dominant terms are perfectly captured by using SINDy, although the magnitudes are slightly different.
These noteworthy results indicate that a governing equation of low-dimensionalized turbulent flows can be obtained from only time series data by using SINDy with appropriate parameter selections.
In other words, we may also be able to construct a machine-learning based reduced-order model for turbulent flows with the interpretable sense as a form of equation, if a well-designed model for mapping into low-dimensional manifolds can be constructed.

\begin{figure}
  \centerline{\includegraphics[clip,width=0.93\linewidth]{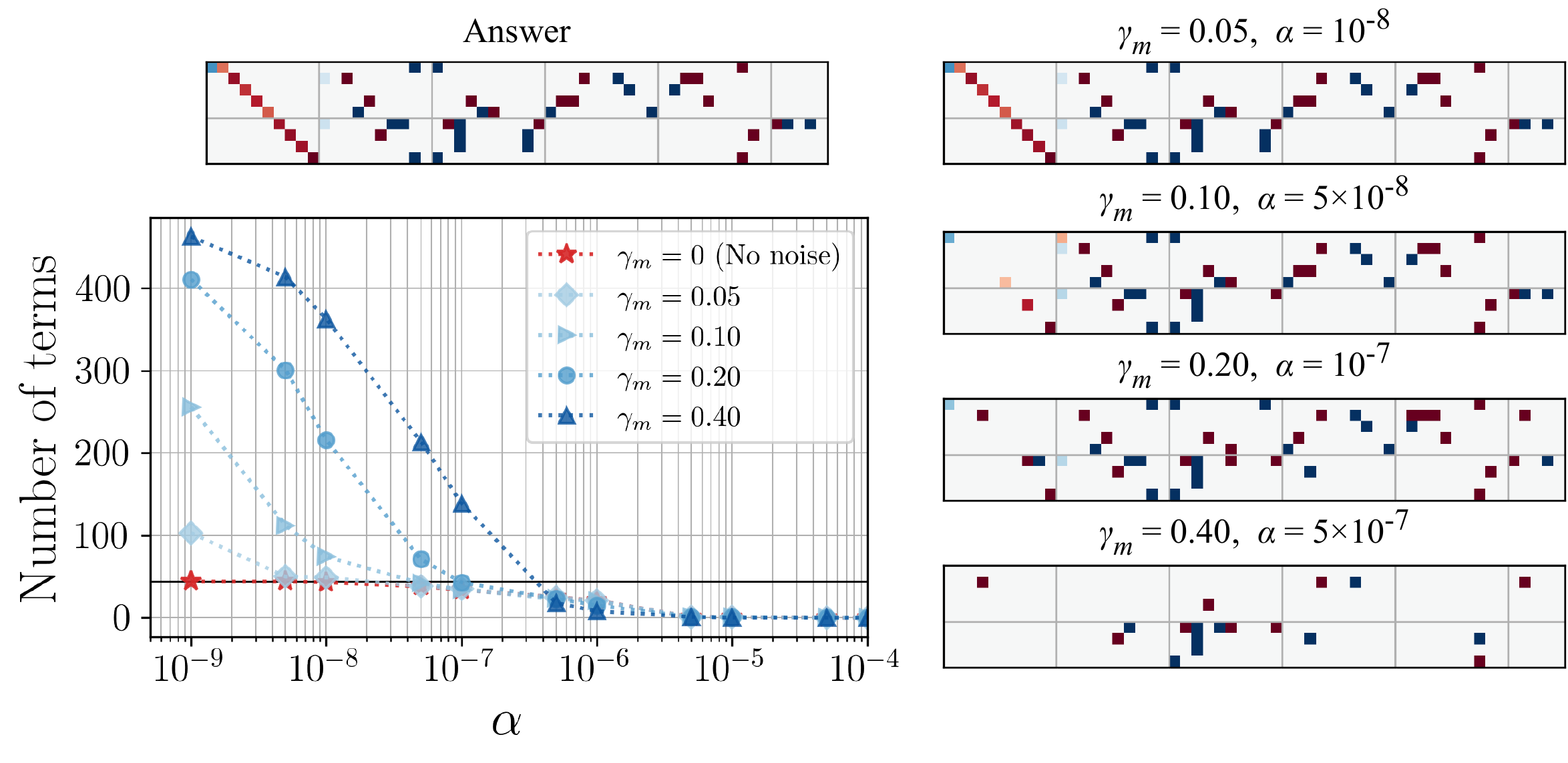}}
  \caption{Noise robustness of SINDy for the nine-shear turbulent flow example.}
  \label{fig_noise}
\end{figure}

The robustness of SINDy against noisy measurements observed in the training pipeline for this example is finally assessed in figure~\ref{fig_noise}. 
We here introduce the Gaussian perturbation for the training data as;
\begin{equation}
    {\bm f}_m = {\bm f} + \gamma_m {\bm n},
\end{equation}
where ${\bm f}$ is a target vector, $\gamma$ is the magnitude of the noise, and ${\bm n}$ denotes the Gaussian noise.
As the magnitude of noise, four cases $\gamma_m=\{0.05, 0.10, 0.20, 0.40\}$ are considered.
With $\gamma = 0.05$, SINDy is still be able to identify the equation accordingly.
However, the first-order terms are eliminated and unnecessary high-order terms are added with increasing the noise magnitude of the training data, although the whole trend of coefficient matrix can be captured.
These results imply that care should be taken for noisy observations in training data depending on problem setting users handle, although the SINDy can guarantee its robustness for a slight perturbation.

\section{Conclusion}
\label{sec:conc}

We performed a sparse identification of nonlinear dynamics (SINDy) for low-dimensionalized fluid flows and investigated influences of the regression methods and parameter considered for construction of SINDy-based modeling.
Following our preliminary test, the SINDys with two regression methods, the TLSA and the Alasso, were applied to the examples of a wake around a cylinder, its transient process, and a wake of two-parallel cylinders, with a convolutional neural network-based autoencoder (CNN-AE).
The CNN-AE was employed to map a high-dimensional flow data into a two-dimensional latent space using nonlinear functions.
Temporal evolution of the latent dynamics could be followed well by using SINDy with an appropriate parameter selection for both examples, although the required number of terms for the coefficient matrix are varied with each other due to the difference in complexity.

At last, we also investigated the applicability of SINDy to turbulence with low-dimensional representation using a nine-shear flow model.
The governing equation could be obtained successfully by utilizing Alasso with the appropriate sparsity constant.
The results indicated that machine-learning based ROM for turbulent flows can be perhaps presented with the interpretable sense as a form of equation, if a well-designed model for mapping high-dimensional complex flows into low-dimensional manifolds can be constructed.
With regard to the low-dimensional manifolds of fluid flows, the novel work by~\citet{Loiseau2018} discussed the nonlinear relationship between lower-order POD modes and higher-order counterparts of cylinder wake dynamics by relating to a triadic interaction among POD modes.
We may be able to borrow the concept there to investigate the possible nonlinear correlations in the data, although it will be tackled in the future since it is tough to apply their idea directly to the nine-equation shear flow model which is composed of more complex relationships among modes.
Otherwise, combination with uncertainty quantification~\citep{MFRFT2020}, transfer learning~\citep{guastoni2020prediction,MFZF2020}, and Koopman-based frameworks~\citep{eivazi2020recurrent} may also be possible extensions toward more practical applications.

The current results for CNN-SINDy modeling make us skeptical against applications of CNN-AE to the flows where a lot of spatial modes are required to represent the energetic field, e.g., turbulence, despite that recent several studies have shown its potential for unsteady laminar flows~\citep{THBSDBDY2019,carlberg2019recovering,bukka2021assessment,xu2020multi,agostini2020exploration}.
Since a few studies have also recently recognized the challenges of the current form of CNN-AE for turbulence~\citep{FNF2020,nakamura2020extension,glaws2020deep,BHT2020}, care should be taken for the choice of low-dimensionalization tools depending on flows users handle.
Hence, the utilization of other well-sophisticated low-dimensionalization tools such as spectral POD~\citep{towne2018spectral,abreu2020spectral}, t-distributed stochastic neighbor embedding~\citep{wu2017visualization}, hierarchical AE~\citep{FNF2020}, and locally linear embedding~\citep{roweis2000nonlinear,ehlert2019locally} can be considered to achieve better mapping of flow fields.
Taking constraints for coordinates extracted via AE is also a candidate to promote the combination of AE modes and conventional flow control tools~\citep{CLKB2019,jayaraman2020data,ladjal2019pca,gelss2019multidimensional}.
We finally believe that the results of our examples above enable us to have a strong motivation for future work and notice the remarkable potential of SINDy for fluid dynamics.

\section*{Acknowledgement}
This work was supported from the Japan Society for the Promotion of Science (KAKENHI grant number: 18H03758, 21H05007).
The authors thank Prof. Shinnosuke Obi, Prof. Keita Ando, Mr. Masaki Morimoto, Mr. Taichi Nakamura, Mr. Shoei Kanehira (Keio Univ.), Mr. Kazuto Hasegawa (Keio Univ., Polimi), and Prof. Kunihiko Taira (UCLA) for fruitful discussions.

\section*{Declaration of interest}

The authors report no conflict of interest.

{
\section*{{Appendix A. Regression methods for SINDy}}

As introduced in section 1, the regression method used to obtain the coefficient matrix $\Xi$ is important since the resultant coefficient matrix $\Xi$ may vary by the choice of regression method.
Here, let us consider a typical regression problem,
\begin{align}
    \bm{P}=\bm{Q} \bm{\beta},
\end{align}
where $\bm{P}$ and $\bm{Q}$ denote the response variables and the explanatory variable, respectively.
In this problem, we aim to obtain the coefficient matrix $\bm{\beta}$ representing the relationships between $\bm{P}$ and $\bm{Q}$.
Using the linear regression method, an optimized coefficient matrix $\bm{\beta}$ can be found by minimizing the error between the left-hand and right-hand side such that
\begin{align}
    \bm{\beta}={\rm argmin} \left\| \bm{P}-\bm{Q}\bm{\beta}  \right\|^2. \label{eq:2.10}
\end{align}
The thresholded least square algorithm (TLSA), originally used in \citet{BPK2016a} for performing SINDy, is based on this formulation in equation (\ref{eq:2.10}) and obtains the sparse coefficient matrix by substituting zero for the coefficient below the threshold $\alpha$.
Its algorithm is summarized in algorithm 1.

\begin{algorithm} 
\begin{algorithmic}[1]
\caption{Thresholded least square algorithm}
\STATE $\alpha \xleftarrow{}$ a certain value (set threshold).
\REPEAT
\STATE Obtain $\bm{\beta}={\rm argmin} \left\| \bm{P}-\bm{Q}\bm{\beta}  \right\|^2$.
\STATE $\beta_i \xleftarrow{} 0$ where $\beta_i<\alpha$.
\STATE Change $\bm{P}$ and $\bm{Q}$ so that the components which were set to zero in the previous step will be output as zero.
\UNTIL{the estimate $\bm{\beta}$ no longer changes}
\end{algorithmic}
\end{algorithm}

Although the linear regression method has widely been used due to its simplicity, \citet{HK1970} cautioned that estimated coefficients are sometimes large in absolute value for the non-orthogonal data. 
This is known as the overfitting problem.
As a considerable surrogate method, they proposed the Ridge regression \citep{HK1970}. 
In this method, the squared value of the components in the coefficient matrix is considered as the constraint of the minimization process such that
\begin{align}
    \bm{\beta} ={\rm argmin} \left\| \bm{P} - \bm{Q}\bm{\beta} \right\|^2 + \alpha \sum_j \beta_j^2,
\end{align}
where $\alpha$ denotes the weight of the constraint term.
For uses of SINDy, the coefficient matrix $\bm{\beta}$ is desired to be sparse, i.e., some components are zero, so as to avoid construction of a complex model and overfitting.
As the sparse regression method, the least absolute shrinkage and selection operator (Lasso) \citep{Tibshirani1996} was proposed. 
With Lasso, the sum of the absolute value of components in the coefficient matrix is incorporated to determine the coefficient matrix:
\begin{align}
    \bm{\beta} ={\rm argmin} \left\| \bm{P} - \bm{Q}\bm{\beta} \right\|^2 + \alpha \sum_j |\beta_j|.
\end{align}
If the sparsity constant $\alpha$ is set to a high value, the estimation error becomes relatively large while the coefficient matrix results in sparse.

The Lasso algorithm introduced above has, however, two drawbacks and there are solutions for each: the elastic net (Enet) and the adaptive Lasso (Alasso), as mentioned in introduction.
Enet \citep{ZH2005} combines $L_1$ and $L_2$ errors, i.e., those of Ridge and Lasso. 
The optimal coefficient $\bm{\beta}$ is obtained as
\begin{align}
    \bm{\beta} = \left( 1+\frac{\alpha_2}{n} \right) \left\{ {\rm argmin} \left\| \bm{P} - \bm{Q}\bm{\beta} \right\|^2 + \alpha_1 \sum_j |\beta_j| + \alpha_2 \sum_j \beta_j^2 \right\}.
\end{align}
The following two parameters are set for the Enet: the sparsity constant $\alpha=\alpha_1+\alpha_2$ and the sparse ratio $\alpha_1 / (\alpha_1 + \alpha_2)$, respectively. 
The property for the sparsity constant $\alpha$ is the same as that of Lasso: a higher $\alpha$ results in large error and sparse estimation. 
The sparse ratio is the ratio of $L_1$ and $L_2$ errors as seen its definition.
The sparse model can be obtained with a high $L_1$ ratio, although the collinearity problem may not be solved.
Another solution, the adaptive Lasso \citep{Zou2006}, uses adaptive weights for the
penalizing coefficients in $L_1$ penalty term. 
In the adaptive Lasso, $\bm{\beta}$ are given by
\begin{align}
    \bm{\beta} ={\rm argmin} \left\| \bm{P} - \bm{Q}\bm{\beta} \right\|^2 + \alpha \sum_j w_j |\beta_j|,
\end{align}
where $w_j=(|\beta_j|)^{-\delta}$ denotes the adaptive weight with $\delta >0$. 
The use of $w_j$ enables us to avoid the issue of multiple local minima.
The weights $w_j$ and coefficients $\bm{\beta}$ are updated repeatedly like the algorithm 2:
\begin{algorithm} 
\begin{algorithmic}[1]
\caption{Adaptive lasso}
\STATE $\alpha \xleftarrow{}$ a certain value (set the sparsity constant).
\STATE $w_j \xleftarrow{} 1$ (initialization).
\REPEAT
\STATE $\bm{Q}^{**} \xleftarrow{} \bm{Q}/w_j$.
\STATE Solve the Lasso problem: $\bm{\beta}^{**} ={\rm argmin} \left\| \bm{P} - \bm{Q}^{**}\bm{\beta} \right\|^2 + \alpha \sum_j w_j |\beta_j|$.
\STATE $\bm{\beta} \xleftarrow{} \bm{\beta}^{**}/w_j$.
\STATE $w_j \xleftarrow{} (|\beta_j|)^{-\delta}$.
\UNTIL{the estimate $\bm{\beta}$ no longer changes}
\end{algorithmic}
\end{algorithm}

In general it is necessary to choose the appropriate value for the threshold or the sparsity constant $\alpha$ in order to obtain a desired equation which is sparse and in an interpretable form.
Hence, we focus on seeking the appropriate $\alpha$ of each regression method in this study.
The algorithm for the parameter search applied to the current study is described in algorithm 3.
\begin{algorithm} 
\begin{algorithmic}[1]
\caption{Parameter search}
\STATE Prepare data $\bm{X}$ and $\dot{\bm{X}}$.
\STATE Prepare the set of sparse parameters $\alpha$.
\FORALL{$\alpha$}
\STATE Perform SINDy to obtain the ordinary differential equation at a certain $\alpha$.
\STATE Assess the candidate model with an evaluation index.
\ENDFOR
\STATE Choose the best model by the assessment.
\STATE Model is identified.
\end{algorithmic}
\end{algorithm}
Regarding the assessment ways used in the procedures 5 and 7 in algorithm 3, number of total remained terms of the obtained equation are utilized as the evaluation index for the preliminary test.
For cylinder examples, the number of total terms, the error in the 
amplitudes of obtained wave forms, and the error ratio in the period between the obtained wave and the solution are assessed.
}

{
\section*{Appendix B. Preliminary tests with low-dimensional problems}
\subsection*{Pre-test 1: van der Pol oscillator}

As a first preliminary test of performing SINDy, we consider the van der Pol oscillator~\citep{van1934nonlinear}, which has a nonlinear damping. 
The governing equations are 
\begin{align}
    \frac{dx}{dt}&=y, \label{eq:vdp1}\\
    \frac{dy}{dt}&=-x+\kappa y -\kappa x^2 y, \label{eq:vdp2}
\end{align}
where $\kappa>0$ is a constant parameter and we set $\kappa=2$ in this preliminary test.
The trajectory converges to a stable limit cycle determined by $\kappa$ under any initial state except for the unstable point ($x=0, y=0$) as shown in figure~\ref{fig_vdp}.
\begin{figure}
  \centerline{\includegraphics[clip,width=0.7\linewidth]{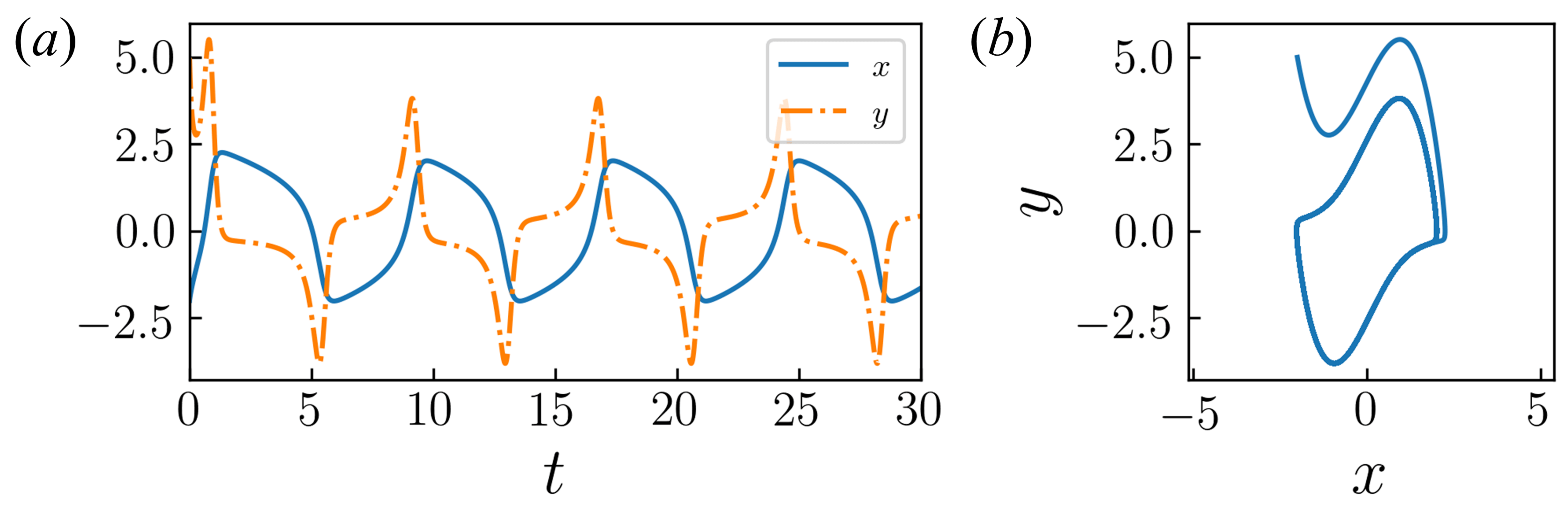}}
  \caption{Dynamics of van der Pol oscillator with $\kappa=2$: ($a$) time history and ($b$) trajectory on $x-y$ plane. The initial point is set to $(x,y)=(-2,5)$ in this example.}
  \label{fig_vdp}
\end{figure}

Let us perform SINDy with each of the four regression methods, i.e., TLSA, Lasso, Enet and Alasso, with various sparsity constant $\alpha$.
Here, we use 20000 discretized points as the training data sampled within $t=[0, 200]$ with the time step $\Delta t=\SI{1e-2}{}$, although we will investigate the dependence of SINDy performance on both the length of time step and the length of training data range later.
The library matrix is arranged including up to the 5th order terms.
In this example, we assess candidate models by using the total number of terms contained in the 
obtained ordinary differential equation. 
As denoted in equations~\ref{eq:vdp1} and~\ref{eq:vdp2}, the true model has four non-zero terms in total in two governing equations.
The relationship between the total number of terms in the candidate models and $\alpha$ is shown in figure~\ref{fig_vdp_4methods}.
\begin{figure}
  \centerline{\includegraphics[clip,width=1\linewidth]{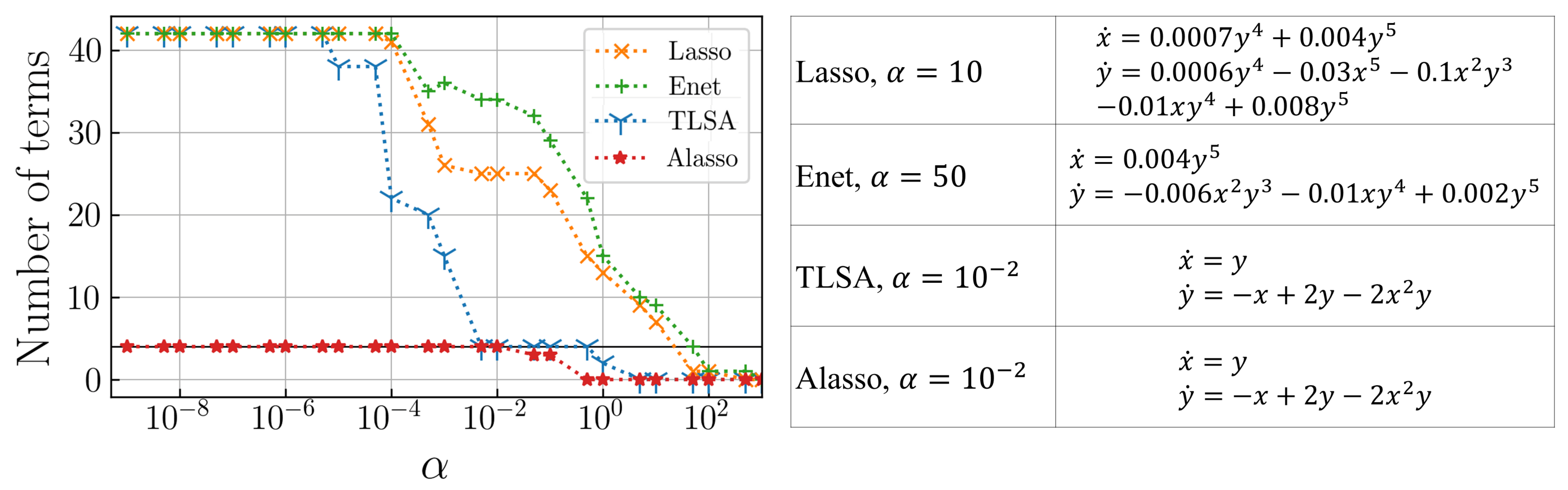}}
  \caption{Relationship between $\alpha$ and the number of terms, and the examples of obtained equations for four regression methods.}
  \label{fig_vdp_4methods}
\end{figure}
As shown, the uses of Lasso and Enet cannot reach the true model whose total number of terms is four.
On the other hand, using TLSA or Alasso, the correct governing equation can be identified.
It is striking that we should choose the appropriate parameter $\alpha$ for the correct model identification, as shown in figure~\ref{fig_vdp_4methods}.
In this sense, the Alasso is suitable for the problem here since a wide range of $\alpha$, i.e., $10^{-9}\leq \alpha \leq 10^{-2}$, can provide the correct equation.
The difference here is likely due to the scheme of each regression.
The TLSA is unable to discard the components of $\beta$ with low $\alpha$ values: in other words, a lot of terms are remained through the iterative process.
On the other hand, the Alasso can identify the dominant terms clearly thanks to adaptive weights even if $\alpha$ is relatively low.

To investigate the dependence on the time step of the input data, SINDy is performed to the data with different time stpdf $\Delta t=\SI{1e-2}{}$, $0.1$ and $0.2$.
Here, we consider only TLSA and Alasso following the aforementioned results.
Note that the number of input data is different due to time step since the integration time length is fixed to $t_{\rm max}=200$ in all cases. 
We have checked in our preliminary tests that this difference has no significant effect on the
results.
\begin{figure}
  \centerline{\includegraphics[clip,width=1\linewidth]{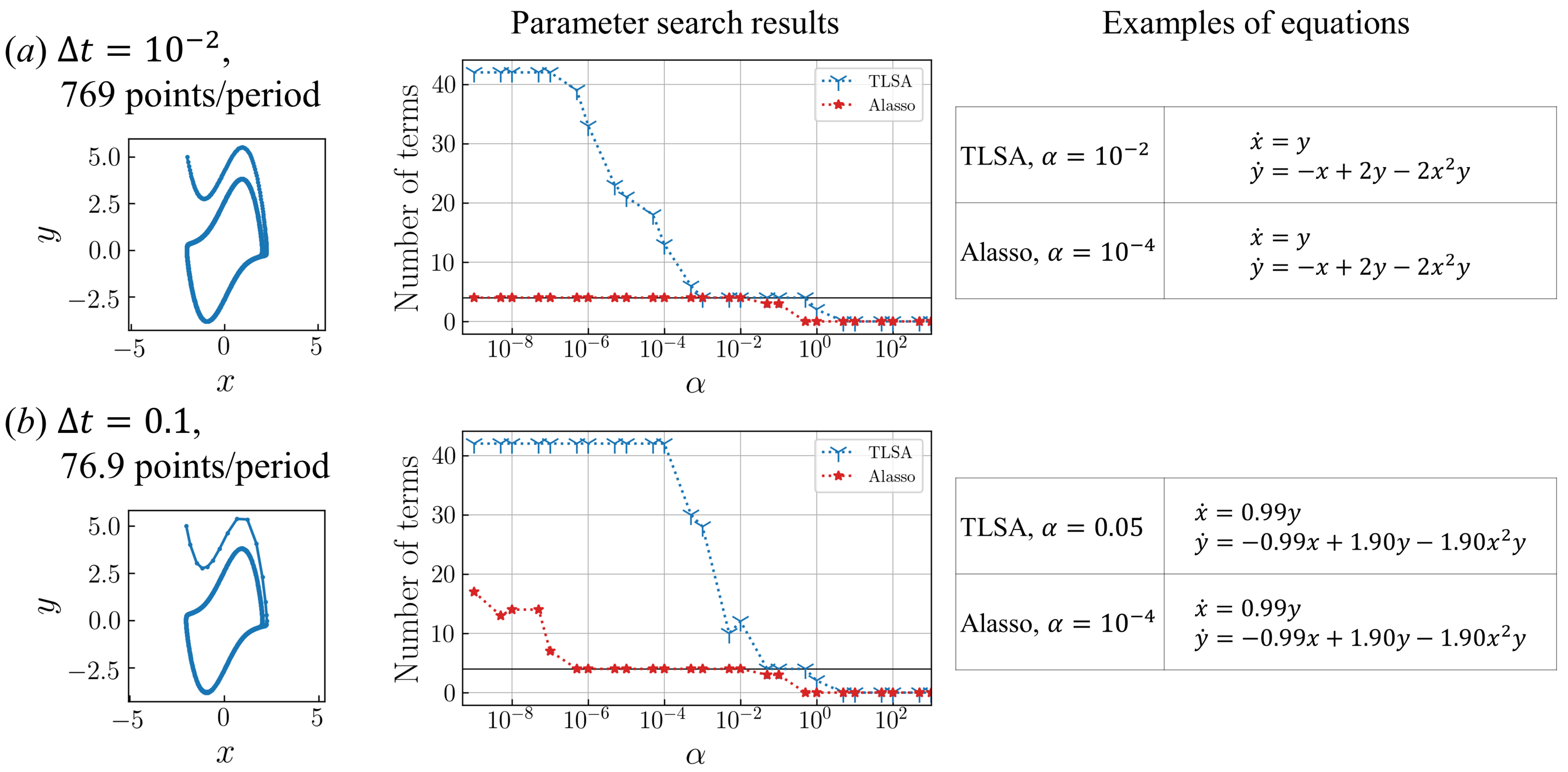}}
  \centerline{\includegraphics[clip,width=1\linewidth]{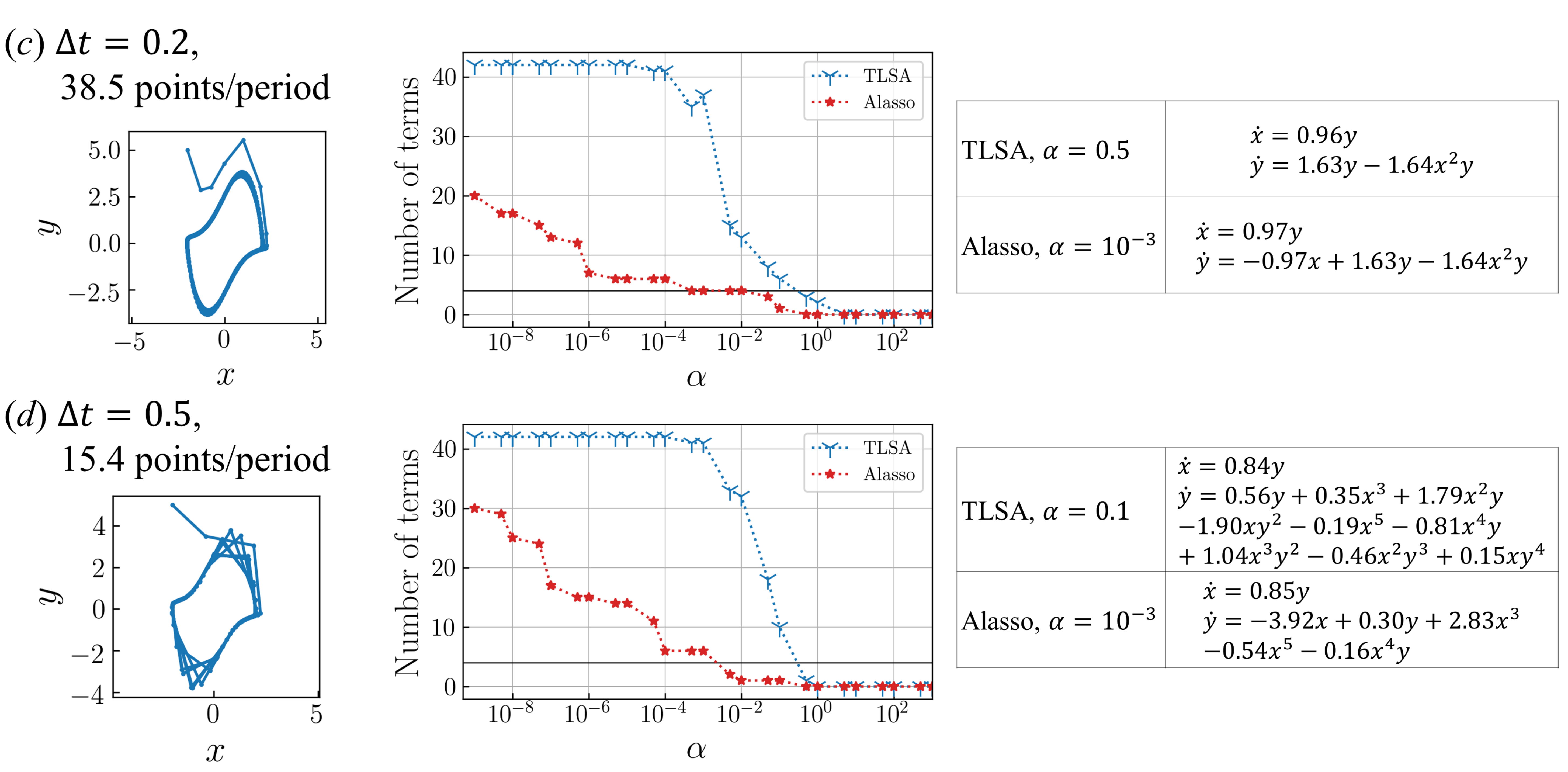}}
  \caption{Trajectory on $x-y$ plane, the relationship between $\alpha$ and number of terms, and the examples of obtained equations for data for different time stpdf: ($a$) $\Delta t =\SI{1e-2}{}$, ($b$) $\Delta t =0.1$, ($c$) $\Delta t =0.2$ and ($d$) $\Delta t=0.5$.}
  \label{fig_vdp_time_step}
\end{figure}
The trajectory, the results for the parameter search, and the example of obtained equations are summarized in figure~\ref{fig_vdp_time_step}.
With the fine data, i.e., $\Delta t=\SI{1e-2}{}$, the governing equations are correctly identified with some $\alpha$ for both methods.
The true model can be found even if we use very small $\alpha$, i.e., $\alpha=\SI{1e-8}{}$ with Alasso as mentioned before.
With increasing $\Delta t$, i.e., $\Delta t=0.1$, the coefficients of terms are slightly underestimated.
Additionally, the range of $\alpha$ where the number of terms is correctly identified as four becomes narrower with both methods.
With a wider time step, i.e., $\Delta t=0.2$, the dominant terms can be found only with Alasso. 
On the other hand, the true model cannot be obtained with further wider time step data, i.e., $\Delta t=0.5$, even with Alasso, due to the temporal coarseness as shown in figure~\ref{fig_vdp_time_step}$(d)$.
It suggests that the data with appropriate time step is necessary for construction of the model.
In sum, Alasso is superior in terms of the ability to correctly identify the dominant terms for a wider range of sparsity constant.

\begin{figure}
  \centerline{\includegraphics[clip,width=0.8\linewidth]{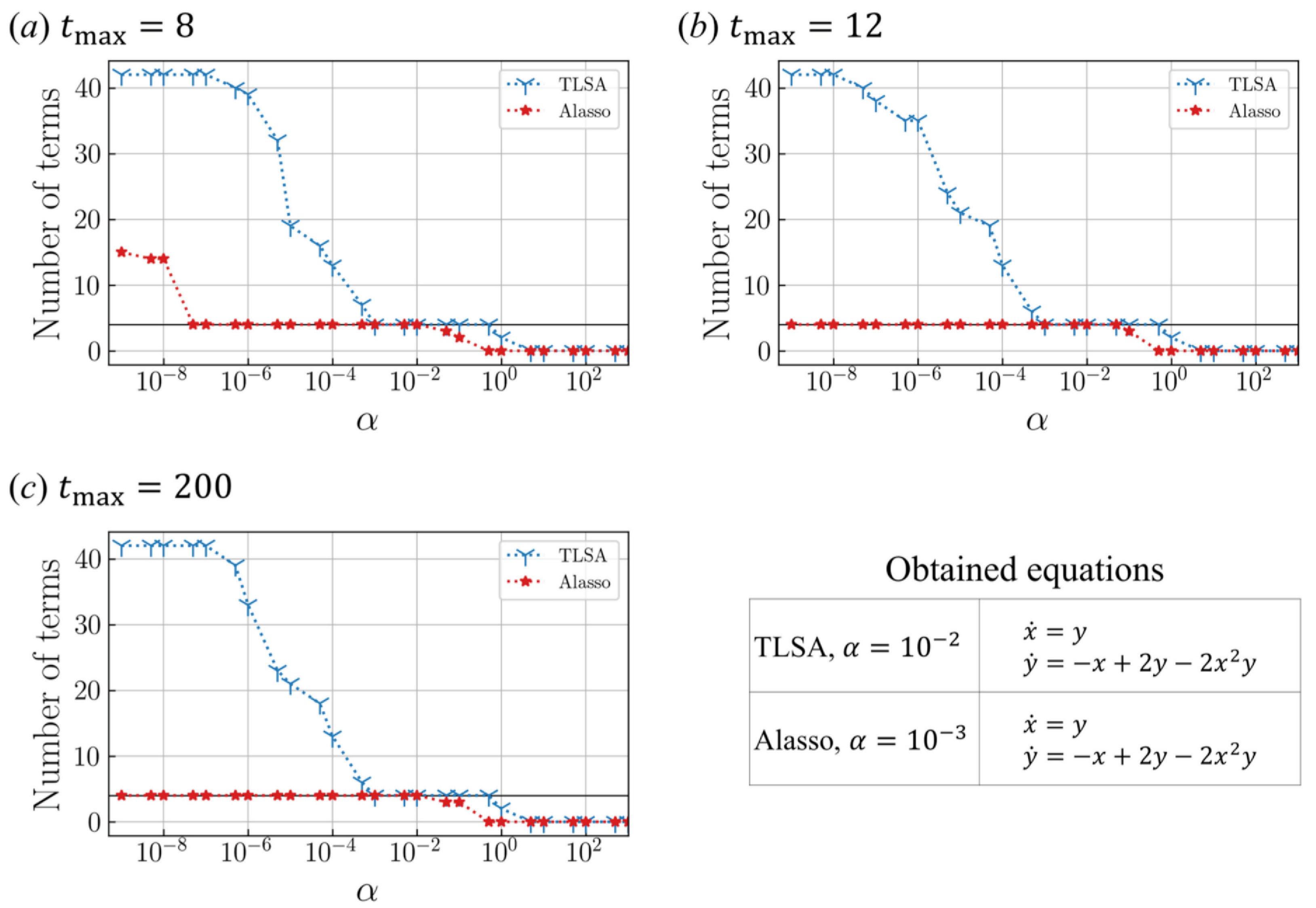}}
  \caption{{The dependence of the performance of SINDy on the length of training data range: $(a)$ $t=[0, 8]$, $(b)$ $t=[0,12]$, and $(c)$ $t=[0,200]$ (baseline).}}
  \label{fig_tdr_vdp}
\end{figure}

Let us then investigate the dependence of the performance on the length of training data range, as demonstrated in figure~\ref{fig_tdr_vdp}.
We here consider three length of training data range as $(a)$ $t=[0, 8]$, $(b)$ $t=[0,12]$, and $(c)$ $t=[0,200]$ (baseline), while keeping the time step $\Delta t=\SI{1e-2}{}$.
As shown, the trends of the relationship between the sparsity constant and the number of terms in the identified equation are similar over the covered time length.
In addition, the SINDys with the TLSA ($\alpha =\SI{1e-2}{}$) and Alasso ($\alpha =\SI{1e-3}{}$) are able to identify the equations successfully for all time length cases.
Only in the case of $t = [0,8]$, the model identification is not performed well with the Alasso around $\alpha =\SI{1e-8}{}$.
The time length has few effects to model identification if the time length is more than one period.

For considering models in which higher order terms may be dominant due to higher nonlinearity, it should be valuable to find the way to select the appropriate potential terms that should be included in the library matrix.
From this view, the dependence on the number of potential terms is investigated using the data with $\Delta t=\SI{1e-2}{}$.
In the case where the maximum order of the potential terms is 5th, the true model can be found with two regression methods, i.e., TLSA and Alasso, as shown in figure~\ref{fig_vdp_time_step}($a$).
Here, the case with higher maximum order, i.e., 15th and 20th, are also investigated as shown in figure~\ref{fig_vdp_pote}.
\begin{figure}
  \centerline{\includegraphics[clip,width=0.9\linewidth]{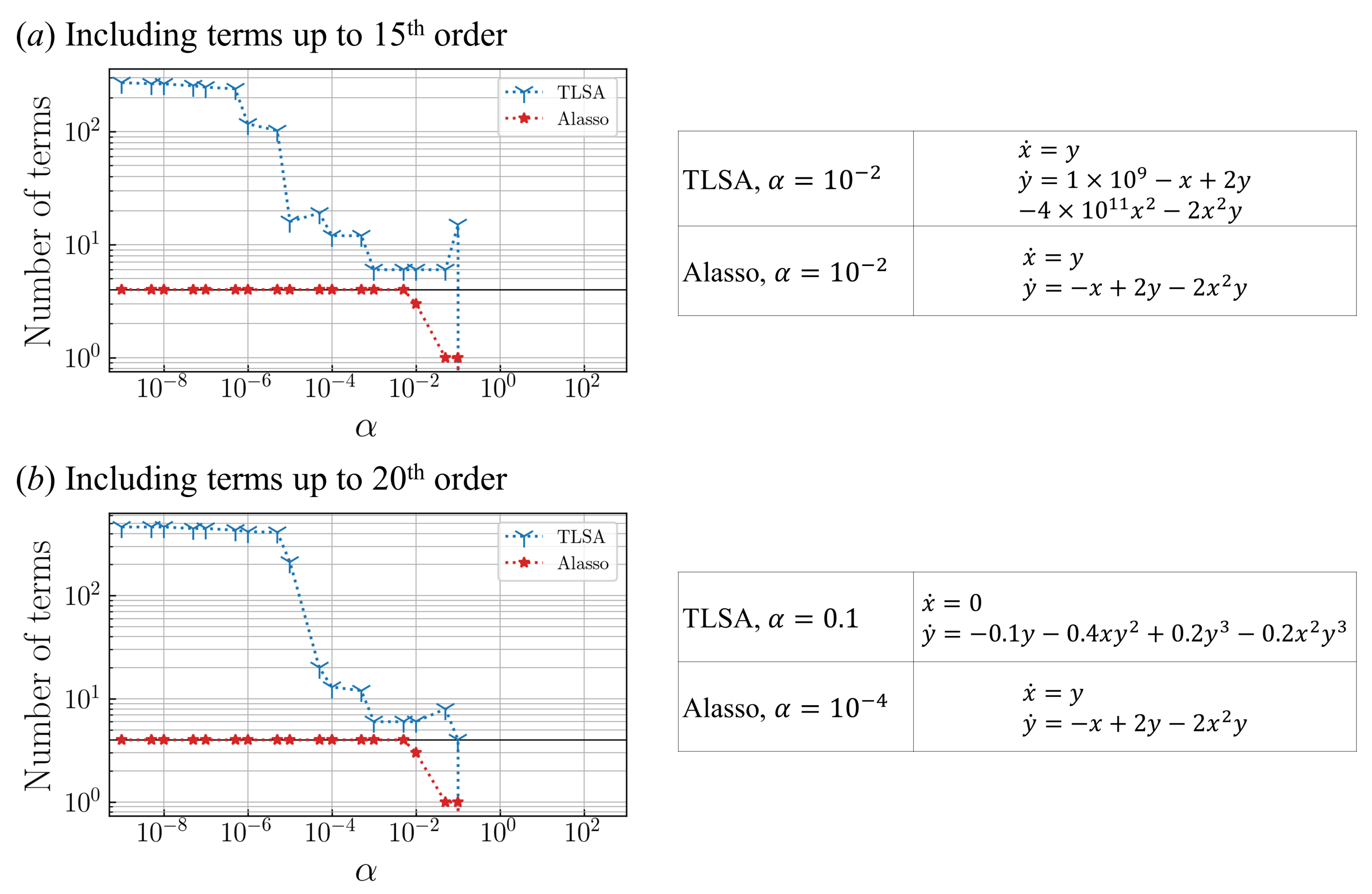}}
  \caption{Relationship between $\alpha$ and number of terms and the examples of obtained equations for data for different library matrices: ($a$) including up to 15th potential terms and ($b$) including up to 20th potential terms.}
  \label{fig_vdp_pote}
\end{figure}
Note that there are 136 and 231 potential terms for each differential equation in these cases.
For both cases with different number of potential terms, we cannot find the true model with TLSA while the true model is identified in a wide range of $\alpha$ with Alasso.
Furthermore, in the case including up to 15th terms, some coefficients in the equation obtained by TLSA become huge.
{These large coefficients are due to the different functions included in the library being almost colinear.}

\subsection*{Pre-test 2: Lorenz attractor}

As the second preliminary test, we consider the problem of identifying a chaotic trajectory called the Lorenz attractor~\citep{lorenz}. 
It is written in a form of nonlinear ordinary differential equations:
\begin{align}
    \frac{dx}{dt}&=-{\sigma}x+{\sigma}y,\\
    \frac{dy}{dt}&={\rho}x-y-xz,\\
    \frac{dz}{dt}&=-{\iota}z+xy.
\end{align}
\citet{lorenz}~showed that the dynamics governed by the equations above shows chaotic behavior with the certain coefficients, i.e., $(\sigma, \rho, \iota)$ = (10, 28, $8/3$) with the initial values $(x,y,z)=(-8,8,27)$, as shown in figure~\ref{fig_lorenz}.
\begin{figure}
  \centerline{\includegraphics[clip,width=0.9\linewidth]{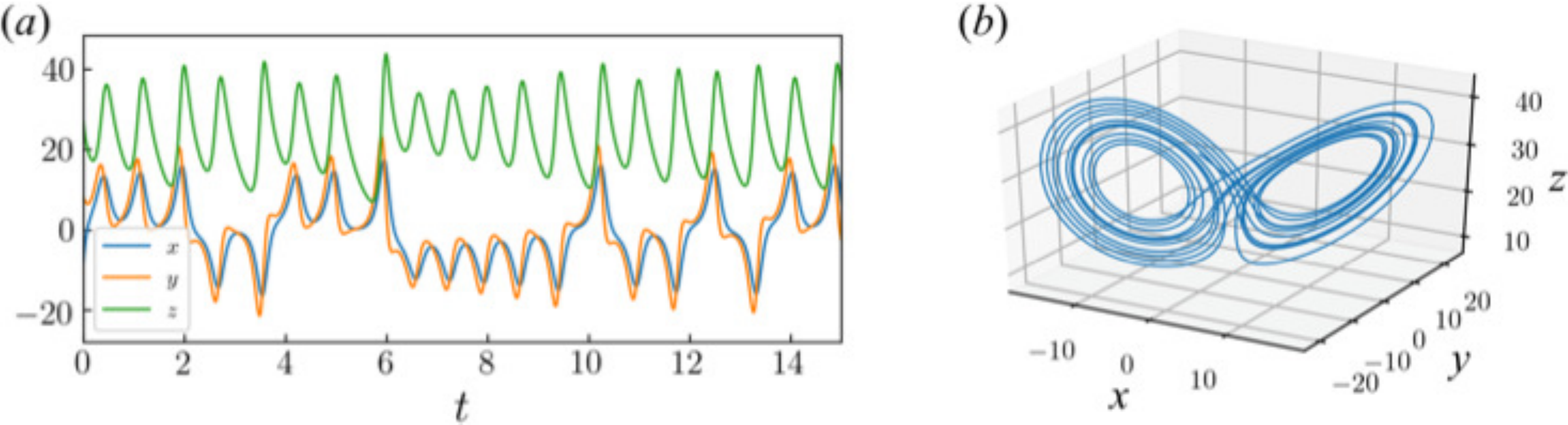}}
  \caption{Dynamics of the Lorenz attractor: $(a)$ time history and $(b)$ trajectory.}
  \label{fig_lorenz}
\end{figure}

As with the preliminary test of van der Pol oscillator, we consider four regression methods: TLSA, Lasso, Enet and Alasso, as shown in figure~\ref{fig_lorenz_4methods}.
Note that 10000 discretized points with $\Delta t=\SI{1e-2}{}$ from the time range of $t=[0,100]$ are used as the training data and the potential terms up to 5th order are considered here.
\begin{figure}
  \centerline{\includegraphics[clip,width=0.9\linewidth]{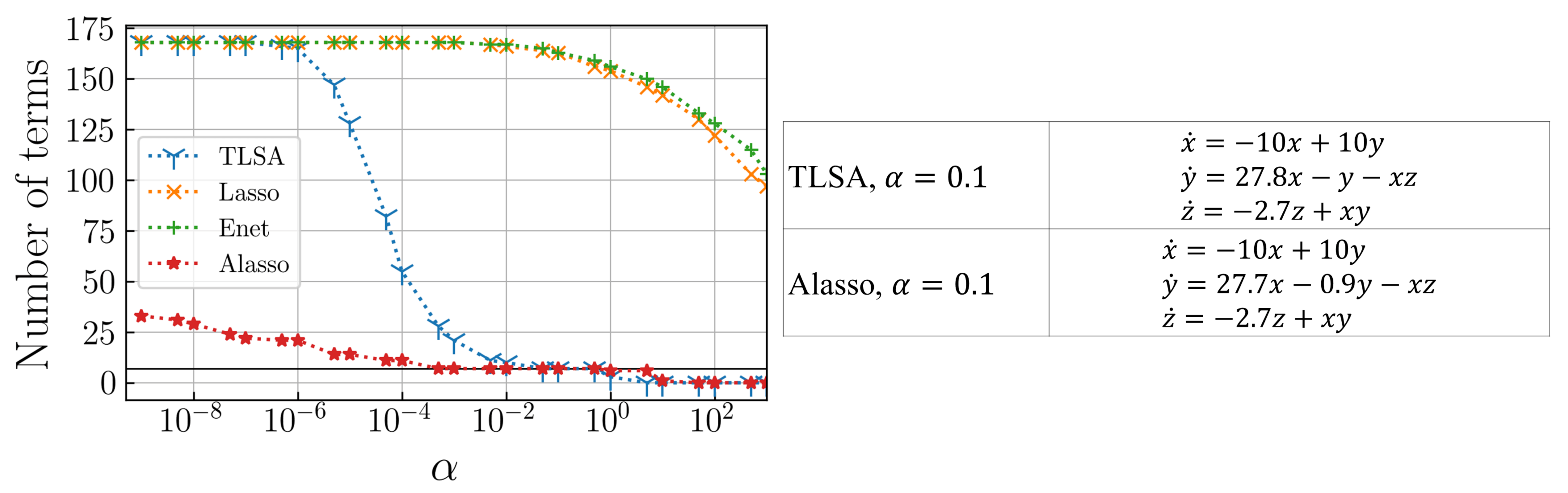}}
  \caption{Relationship between $\alpha$ and the total number of terms, with the examples of obtained equations.}
  \label{fig_lorenz_4methods}
\end{figure}
Similar to the van der Pol oscillator, we can obtain the true model whose total number of terms is seven with TLSA or Alasso, although Lasso or Enet cannot identify the equation correctly.
Comparing the two methods which can obtain the correct model, the range of $\alpha$, where the governing equations are identified, is wider using Alasso.
\begin{figure}
  \centerline{\includegraphics[clip,width=1\linewidth]{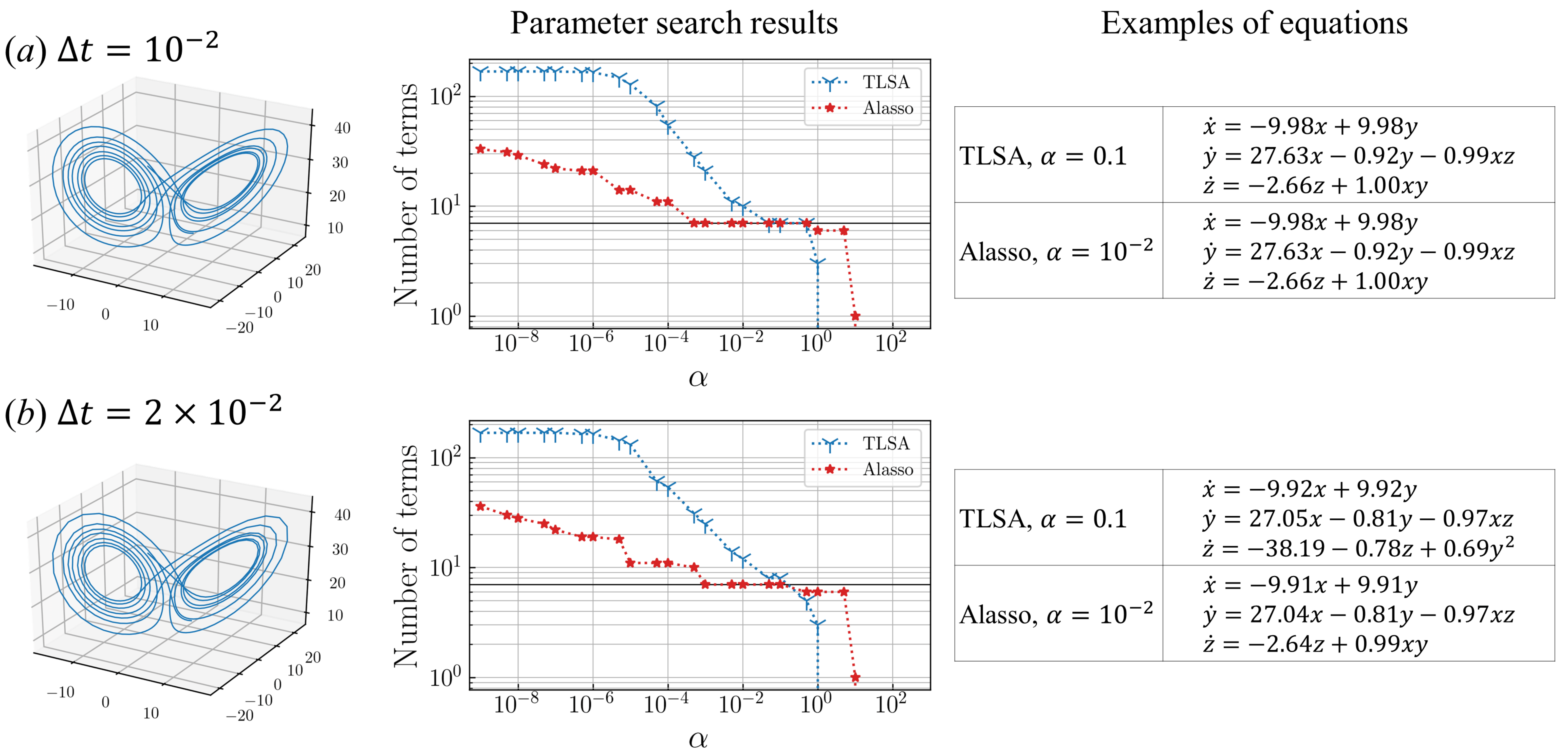}}
  \caption{Trajectory on $x-y$ plane, the relationship between $\alpha$ and number of terms, and the examples of obtained equations for data for different time stpdf: ($a$) $\Delta t =\SI{1e-2}{}$ and ($b$) $\Delta t =\SI{2e-2}{}$.}
  \label{fig_lorenz_time_step}
\end{figure}
We also perform SINDy with different time stpdf as shown in figure~\ref{fig_lorenz_time_step}.
Note that the considered time range is not changed depending on the time stpdf: in other words, the number of training data is changed per a considered time step, which is the same setting as that for the van der Pol example.

\begin{figure}
  \centerline{\includegraphics[clip,width=0.9\linewidth]{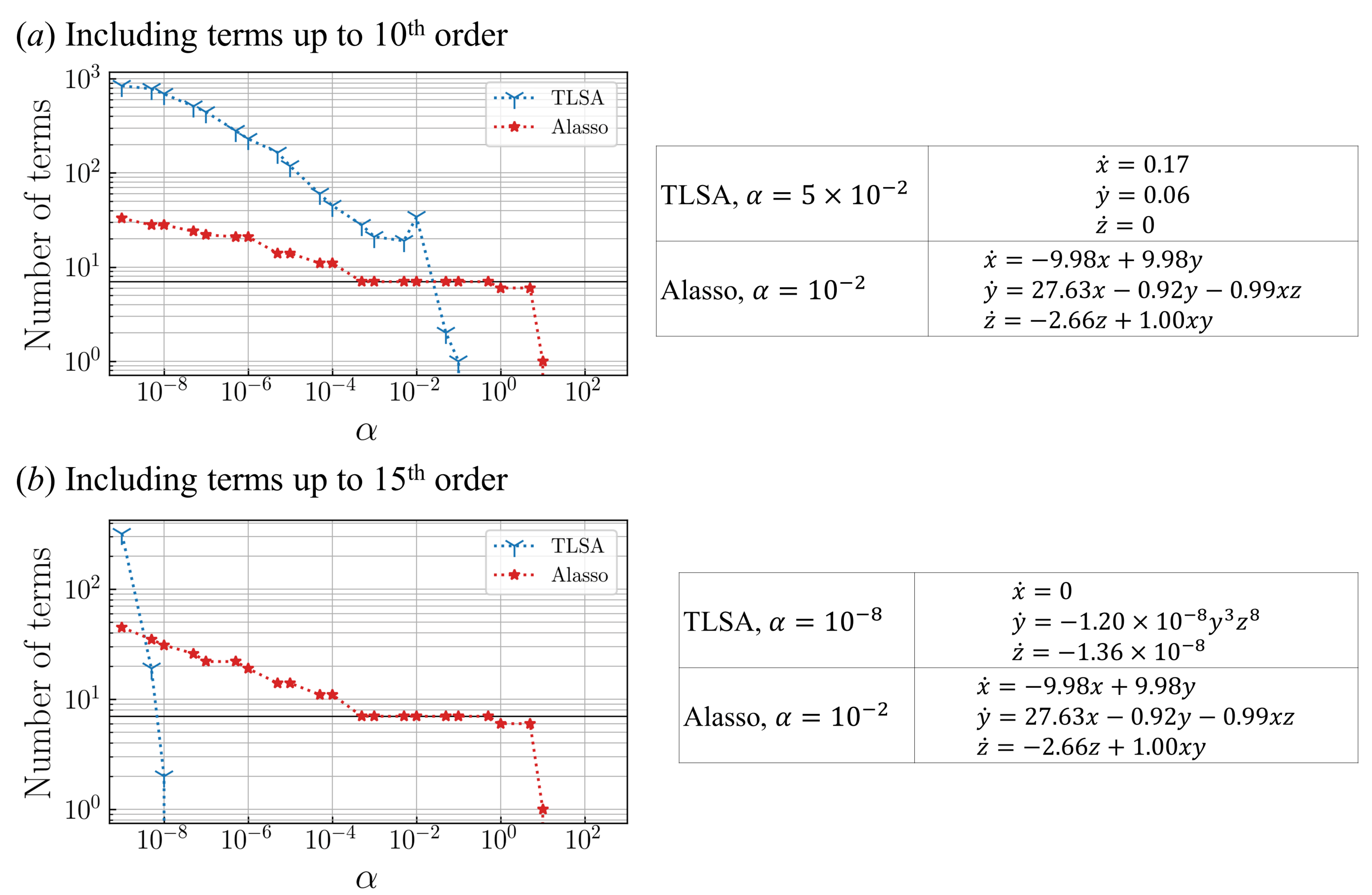}}
  \caption{Relationship between $\alpha$ and number of terms, and the examples of obtained equations for data for different library matrices: ($a$) including up to 10th potential terms and ($b$) including up to 15th potential terms.}
  \label{fig_lorenz_pote}
\end{figure}
Similarly to the case of van der Pol oscillator, we cannot reach the correct model using TLSA 
unless $\alpha$ is fine-tuned, and the coefficients are underestimated using Alasso with the wider time step data, i.e., $\Delta t=\SI{2e-2}{}$.
Moreover, the dependence on how many orders are considered in the library matrix is investigated in figure~\ref{fig_lorenz_pote}.
There are 286 or 816 potential terms for each differential equation with the case including up to 10th or 15th order terms, respectively.
Analogous to the van der Pol oscillator example, the true model cannot be identified with TLSA; however, we can identify the true equation using Alasso even if there are as many as 816 potential terms with an appropriate range of $\alpha$.
Based on these results, TLSA and Alasso are selected as candidate regressions for high-dimensional complex flow problems.

In addition, we here note that the use of a simple data processing (e.g., the standardization and the normalization) for both AE and SINDy pipelines may help Lasso and Enet to improve the regression ability, although it highly relies on a problem setting.
Taking the data processing usually limits the model for practical uses because these data processing can only be successfully employed when we can guarantee an {\it ergodicity}, i.e., the statistical features of training data and test data are common with each other~\citep{guastoni2020convolutional,MFZF2020}.
When we perform the data processing, an inverse transformation error through the normalization or standardization is usually added since parameters for the processing, e.g., minimum value, max value, average value, and variance, may be changed over the training and test phases.
Hence, toward practical applications, it is better to prepare the regression method which is robust for the amplitudes of data oscillations. 

}

\bibliographystyle{jfm}
\bibliography{jfm}

\end{document}